\documentclass[
twocolumn,
showpacs,preprintnumbers,
superscriptaddress,
showkeys,
amsmath,amssymb,
aps,prl
]{revtex4-2}

\usepackage{graphicx}
\usepackage{dcolumn}
\usepackage{bm}
\usepackage{hyperref}
\usepackage{color,soul}
\usepackage{upgreek}
\DeclareUnicodeCharacter{2212}{\ensuremath{-}}
\begin{document}


\title{The Role of Self-Torques in Transition Metal Dichalcogenide/Ferromagnet Bilayers}

\author{Jan Hidding}
 \email{jan.hidding@rug.nl}
 \affiliation{Zernike Institute for Advanced Materials, University of Groningen, 9747 AG Groningen, The Netherlands}
 
\author{Klaiv M\"ertiri}
 \affiliation{Zernike Institute for Advanced Materials, University of Groningen, 9747 AG Groningen, The Netherlands}
 
\author{Fauzia Mujid}%
 \affiliation{Department of Chemistry, Pritzker School of Molecular Engineering, and James Franck Institute, University of Chicago, Chicago, Illinois 60637, United States}

 \author{Ce Liang}%
 \affiliation{Department of Chemistry, Pritzker School of Molecular Engineering, and James Franck Institute, University of Chicago, Chicago, Illinois 60637, United States}
 
\author{Jiwoong Park}%
 \affiliation{Department of Chemistry, Pritzker School of Molecular Engineering, and James Franck Institute, University of Chicago, Chicago, Illinois 60637, United States}
 
\author{Marcos H. D. Guimar\~{a}es}%
 \email{m.h.guimaraes@rug.nl}
\affiliation{Zernike Institute for Advanced Materials, University of Groningen, 9747 AG Groningen, The Netherlands}

\date{\today}

\begin{abstract}
In recent years, transition metal dichalcogenides (TMDs) have been extensively studied for their efficient spin-orbit torque generation in TMD/ferromagnetic bilayers, owing to their large spin-orbit coupling, large variety of crystal symmetries, and pristine interfaces.
Although the TMD layer was considered essential for the generation of the observed SOTs, recent reports show the presence of a self-torque in single-layer ferromagnetic devices with magnitudes comparable to TMD/ferromagnetic devices.
Here, we perform second-harmonic Hall SOT measurements on metal-organic chemical vapor deposition (MOCVD) grown MoS\textsubscript{2}/permalloy/Al\textsubscript{2}O\textsubscript{3} devices and compare them to a single-layer permalloy/Al\textsubscript{2}O\textsubscript{3} device to accurately disentangle the role of self-torques, arising from the ferromagnetic layer, from contributions from the TMD layer in these bilayers.
We report a damping-like self-torque of
$\sigma_{DL}= \left(-2.5 \pm 0.6 \right) \times10^{5} \frac{\hbar}{2e} (\Omega\cdot \mathrm{m})^{-1}$
in our single-layer permalloy/Al\textsubscript{2}O\textsubscript{3} device, while we observe a weaker $\sigma_{DL}$ with opposite sign in one MoS\textsubscript{2}/permalloy/Al\textsubscript{2}O\textsubscript{3} device, and no significant $\sigma_{DL}$ for all other MoS\textsubscript{2}/permalloy/Al\textsubscript{2}O\textsubscript{3} devices.
The opposite sign of these torques indicates a competition between the self-torque and the torque arising from the TMD layer, which would reduce the observed torque in these bilayers.
In addition, we find a field-like spin-torque conductivity of $\sigma_{FL}=(-2.8 \pm 0.3) \times10^{3} \frac{\hbar}{2e} (\Omega\cdot \mathrm{m})^{-1}$ in a single-layer permalloy/Al\textsubscript{2}O\textsubscript{3} device, which is comparable to control MoS\textsubscript{2}/permalloy/Al\textsubscript{2}O\textsubscript{3} devices and previous reports on similar TMD/FM bilayers, indicating only a minor role of the MoS\textsubscript{2} layer.
Finally, we find a linear dependence of the SOT conductivity on the Hall bar leg/channel width ratio of our devices, indicating that the Hall bar dimensions are of significant importance for the reported SOT strength.
Our results accentuate the importance of delicate details, like device asymmetry, Hall bar dimensions, and self-torque generation, for the correct disentanglement of the microscopic origins underlying the SOTs, essential for future energy-efficient spintronic applications.
\end{abstract}

\maketitle

\section{Introduction}\label{sec:introduction}
Manipulating the magnetization of a magnetic layer by means of a charge current holds an immense promise for more energy-efficient ways of storing and writing information \cite{Manchon2015c, Manchon2019, Dieny2020b}.
By first converting a charge current into a spin current in materials with large spin-orbit coupling (SOC), the spin current can subsequently exert a torque on the magnetization of an interfaced magnetic material \cite{Manchon2019}.
These current-induced torques, originating from the spin-orbit interaction, are referred to as spin-orbit torques (SOTs).
To maximize the SOT strength, an efficient charge-to-spin conversion is advantageous, and thus, materials with large spin-orbit coupling (Pt \cite{Garello2013d, Fan2014, Nguyen2016a}, Pd \cite{Jamali2013, Ghosh2017b, Lee2014}, W \cite{Pai2012, Demasius2016}, Ta \cite{Liu2012c, Avci2014b, Kim2013}, Hf \cite{Torrejon2014, Akyol2016, Ramaswamy2016}, etc.) have been extensively studied \cite{Tanaka2008}.
Two main mechanisms for the charge-to-spin conversion in these materials are the spin-Hall (SHE) and the Rashba-Edelstein effect (REE) \cite{Manchon2019, Shao2021}.
For these polycrystalline spin Hall metals with inversion symmetry, these effects, however, do not possess the ideal symmetry for field-free switching of the magnetization of magnetic layers with perpendicular magnetic anisotropy (PMA) used in modern high-density memory storage \cite{Yu2014b, Liu2020b}.
Therefore, other materials such as topological insulators and two-dimensional (2D) van der Waals crystals have been employed to search for new materials which do allow for field-free switching \cite{Sousa2020, Shao2021, Krizakova2022, Yang2022, Kurebayashi2022}.

In this regard, the family of 2D van der Waals materials called the transition metal dichalcogenides (TMDs) have gained much interest as spin source material, owing to their large SOC, atomically-flat surfaces and broad range of crystal symmetries \cite{Hidding2020b, Husain2020d, Liu2020b}.
The more conventional and well known semiconducting TMDs, such as WSe\textsubscript{2} \cite{Hidding2021c, Shao2016e, Novakov2021c}, WS\textsubscript{2} \cite{Lv2018a}, MoS\textsubscript{2} \cite{Shao2016e, Zhang2016a}, were studied first, due to their air stability and developed wafer-scale growth.
More recently, however, the low symmetry TMDs have gained much interest since the observation of the out-of-plane damping-like torque in TMD/FM bilayers, which is optimal for switching magnets with PMA \cite{MacNeill2017a, Macneill2017c, Guimaraes2018, Stiehl2019d, Stiehl2019c,Shi2019a, Kao2022}.

Previous reports on SOTs in TMD/ferromagnetic (FM) bilayers often consider the TMD as essential for the generation of the observed SOTs, either through bulk effects, such as the spin-Hall effect, or effects arising from the TMD/Py interface, such as the Rashba-Edelstein effect, spin-orbit filtering or spin-orbit precession \cite{Amin2020, Veneri2022}.
More recent reports, however, indicate the presence of a self-torque in single-layer ferromagnetic devices without the presence of a spin source material.
A recent study performed magneto-optic Kerr effect (MOKE) measurements to probe the SOT at the surface of a single-layer Py device and observe a sizable SOT at the Py interface, which is ascribed to a SHE in the Py \cite{Wang2019a}.
Also, electrical measurements on Py capped with SiO\textsubscript{2} or Al\textsubscript{2}O\textsubscript{3} show the presence of field-like and damping-like torques in Py devices \cite{Seki2021a}.
And in metallic bilayers, it was shown that self-induced torques lead to errors in the estimation of the spin-torque strength \cite{Aoki2022}.
These self-torques in ferromagnetic materials make it difficult to accurately determine to what extent the TMD layer is contributing to the SOT \cite{Seki2021a}.

Here, we report second-harmonic Hall measurements on MoS\textsubscript{2}/Py/Al\textsubscript{2}O\textsubscript{3} and single-layer Py/Al\textsubscript{2}O\textsubscript{3} devices to disentangle the contribution of self-torques from the FM layer, and more accurately determine the effect of the MoS\textsubscript{2} layer.
We find that the damping-like self-torque in a Py/Al\textsubscript{2}O\textsubscript{3} device is of opposite sign to the damping-like torque found in one of our MoS\textsubscript{2}/Py/Al\textsubscript{2}O\textsubscript{3} devices, indicating a competition between the two.
Furthermore, we show that the field-like self-torque in a single-layer Py/Al\textsubscript{2}O\textsubscript{3} device can be of similar magnitude compared to MoS\textsubscript{2}/Py/Al\textsubscript{2}O\textsubscript{3} devices, indicating a minor effect of the TMD.
In addition, we study the dependence of the spin-orbit torque on the contact/channel width ratio and find a linear dependence of the field-like spin-torque conductivity on the channel width.
These results indicate the importance of single-layer reference samples and the device specifics for an accurate determination of the microscopic mechanisms underlying the spin-orbit torques.

\begin{figure*}[t]
	\includegraphics{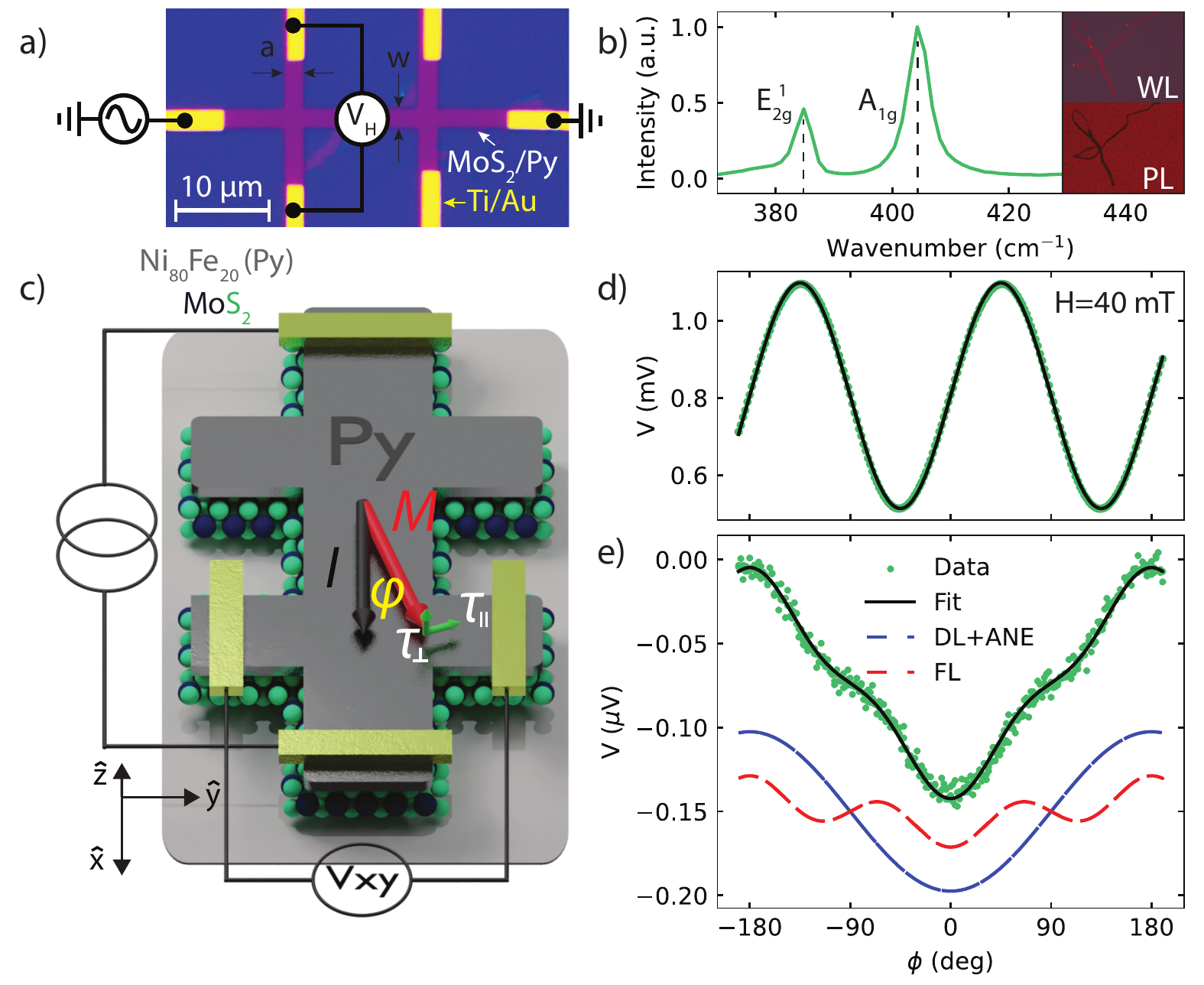}
	\caption{(a) An optical micrograph of an actual MOCVD MoS\textsubscript{2}/Py/Al\textsubscript{2}O\textsubscript{3} device. (b) A Raman spectrum of the MOCVD grown MoS\textsubscript{2} showing the characteristic $E_{2g}^{1}$ and $A_{1g}$ modes of MoS\textsubscript{2}. The two insets on the right depict a white light (WL) and photoluminescence (PL) micrograph of a scratch in the MOCVD grown MoS\textsubscript{2} layer, indicating a strong PL from the monolayer MoS\textsubscript{2}. (c) A schematic of a SOT device with the measurement geometry schematically depicted. A low frequency AC current $I$ (black arrow) is applied through the channel and the first and second-harmonic Hall voltage ($V_{xy}^{\omega(2\omega)}$) are simultaneously measured while the magnetization of the permalloy $M$ (red arrow) is rotated in-plane by an external magnetic field. The current induced in-plane (damping-like) and out-of-plane (field-like) SOTs are depicted with the green arrows $\tau_{\parallel}$ and $\tau_{\perp}$, respectively. (d) The measured first-harmonic and (e) second-harmonic Hall voltage versus in-plane angle of the applied magnetic field (40 mT). (d) A clear $\cos(2\phi)$ dependence is observed due to the planar Hall effect of the Py. (e) The second-harmonic Hall voltage (blue points) is fitted (black line) using Eq.(\ref{eq:SHH}). The dashed blue and red line indicate the separate $\cos(\phi)\cos(2\phi)$ and $\cos(\phi)$ components from Eq.(\ref{eq:SHH}), related to the field-like and damping-like torque, respectively.}
	\label{fig:figure1}
\end{figure*}

\section{Results and Discussion}\label{sec:results}

\subsection{MoS\textsubscript{2}/Py/Al\textsubscript{2}O\textsubscript{3} devices}
We use wafer-scale grown MoS\textsubscript{2} obtained by metal-organic chemical vapor deposition (MOCVD) \cite{Kang2015}.
The MOCVD grown MoS\textsubscript{2} layer is characterized using photoluminescence (PL) microscopy and Raman spectroscopy (see Fig. \ref{fig:figure1}(b)) before device fabrication. 
The two characteristic bands of monolayer MoS\textsubscript{2} at 385 cm$^{-1}$ and 405 cm$^{-1}$, corresponding to the in-plane ($E_{2g}^{1}$) and out-of-plane phonon mode ($A_{1g}$), respectively, are clearly observed, as indicated in Fig. \ref{fig:figure1}b \cite{Li2012}.
Furthermore, a strong and homogeneous PL is obtained using PL microscopy shown in the insets of Fig. \ref{fig:figure1}(b), indicating the homogeneous coverage of monolayer MoS\textsubscript{2} with little strain on the substrate.

\begin{figure*}[t]
	\includegraphics{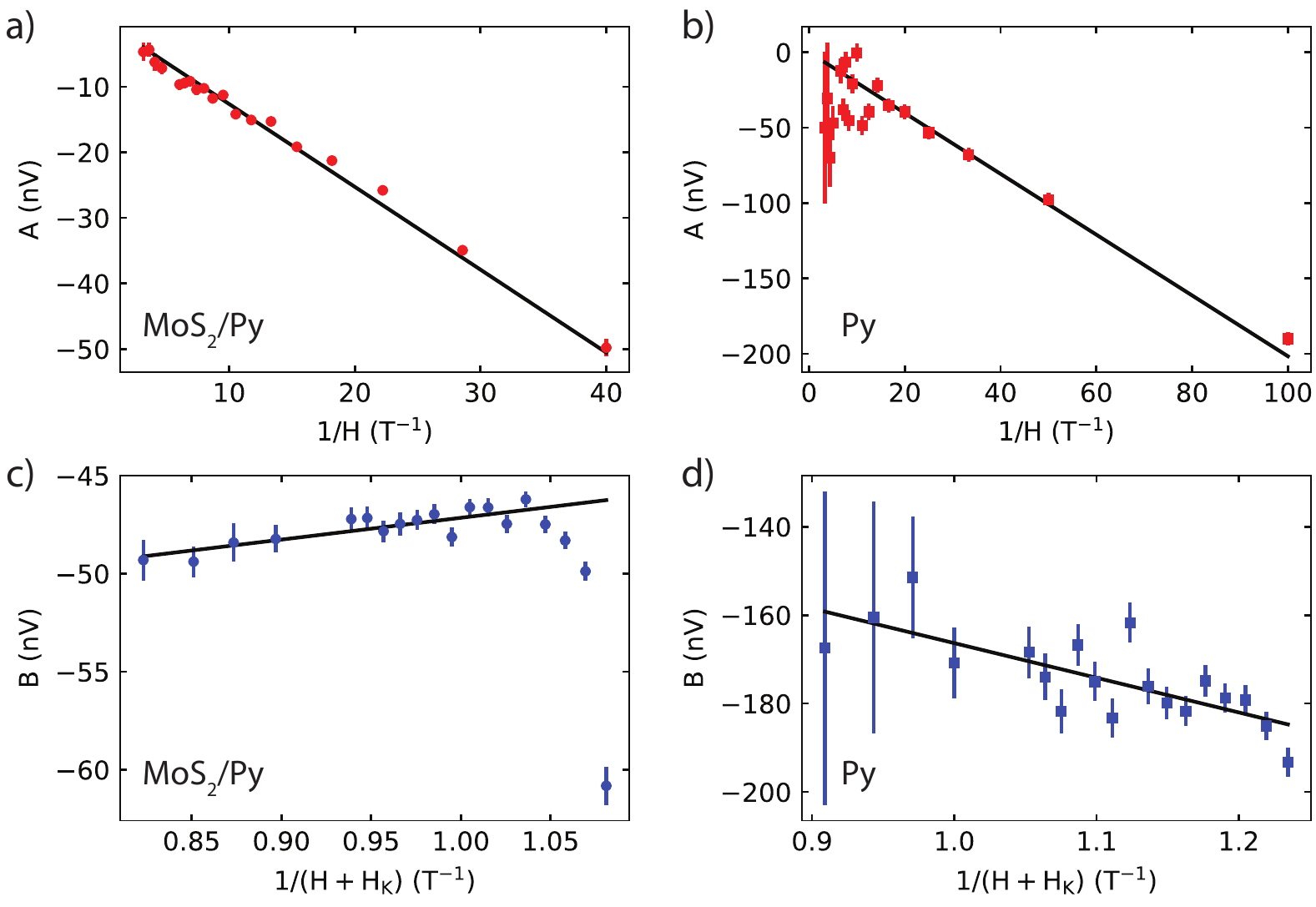}
	\caption{The A- and B-components from Eq.(\ref{eq:SHH}) versus the inverse magnetic field for a MoS\textsubscript{2}/Py/Al\textsubscript{2}O\textsubscript{3}, (a) and (c), and single-layer Py/Al\textsubscript{2}O\textsubscript{3} device, (b) and (d), respectively. The components were extracted by fitting the second-harmonic Hall voltage (as depicted in Fig. \ref{fig:figure1}(e)) to Eq.(\ref{eq:SHH}) for multiple external magnetic field strengths. A clear linear dependence is observed for the A-component in both devices, and are fitted to Eq.(\ref{eq:A}) to extract the field-like torque $\tau_{FL}$. Especially for the MoS\textsubscript{2}/Py/Al\textsubscript{2}O\textsubscript{3} device, the B-component deviates from the linear trend at low magnetic fields, which could be due to electron-magnon scattering. For the MoS\textsubscript{2}/Py/Al\textsubscript{2}O\textsubscript{3} devices, we therefore neglected the fields at low fields (10 mT, 20 mT, and 30 mT) to obtain a better linear fit. Furthermore, we  corrected our data for a systematic offset of 15 mT in the applied field.}
	\label{fig:figure2}
\end{figure*}

Next, multiple MoS\textsubscript{2}/Py/Al\textsubscript{2}O\textsubscript{3} Hall bar devices with different channel width (w)/leg width (a) ratios (Fig. \ref{fig:figure1}a) were fabricated to perform the harmonic Hall measurements.
Here, the first ($V_{xy}^{\omega}$) and second ($V_{xy}^{2\omega}$) harmonic Hall voltage are measured while an applied magnetic field ($H$) is rotated in-plane, making an angle $\phi$ with respect to the current direction (Fig. \ref{fig:figure1}(c), details are described in the Methods section).
When assuming a small magnetic anisotropy compared to $\mu_{0}H$, the magnetization is aligned with the external magnetic field and the first-harmonic Hall voltage ($V_{xy}^{\omega}$) is given by \cite{Garello2013d, Hayashi2014b, Nguyen2021a}:

\begin{equation}\label{eq:FHH}
    V_{xy}^{\omega}(\phi) = I_{0}R_{PHE}\sin(2\phi) + I_{0}R_{AHE}\cos(\theta),
\end{equation}

\noindent where $\theta$ is the magnetic field's polar angle ($\theta=90^{\circ}$ for in-plane measurements) and the $R_{PHE}$ and $R_{A HE}$ are the planar Hall and anomalous Hall effect resistance, respectively.
The first-harmonic Hall voltage ($V_{xy}^{\omega}$), depicted in Fig. \ref{fig:figure1}(d) for a magnetic field of 40 mT, nicely follows a $\sin(2\phi)$ behavior due to the planar Hall effect of the Py layer.
By fitting the data to Eq.(\ref{eq:FHH}) we extract a planar Hall resistance of $R_{PHE} = 0.40 \pm 0.03$ $\Omega$.
In previous reports on exfoliated TMD/FM bilayers, large deviations from the $\sin(2\phi)$ dependence were observed at low magnetic fields, indicating that a strong uniaxial magnetic anisotropy was induced in the Py \cite{Hidding2021c, MacNeill2017a, Stiehl2019c, Stiehl2019d}.
In these reports, the induced anisotropy was attributed to a strong interaction between the Py and the crystalline structure of the underlying TMD.
As the MOCVD grown TMD in our devices has a grain size of around 1 $\upmu$m, our Hall bar covers multiple domains.
Therefore, no induced magnetic anisotropy in the Py from the TMD crystal structure is expected.
This is in line with our observation, as only minor deviations from the $\sin(2\phi)$ fit are observed. 
For devices with a larger Hall bar channel width/leg width ratio, these minor deviations disappear completely, indicating that the minor deviations observed for narrow Hall bars are likely due to shape anisotropy of the Hall bar.

To determine the field-like ($\tau_{FL}$) and damping-like torques ($\tau_{DL}$), the second-harmonic Hall (SHH) voltage is measured (Fig. \ref{fig:figure1}(e)).
The second-harmonic Hall voltage can be described by \cite{Garello2013d, Hayashi2014b}:

\begin{equation}\label{eq:SHH}
    V_{xy}^{2\omega}(\phi) = A\cos(2\phi)\cos(\phi) + B\cos(\phi),
\end{equation}

\noindent where the A- and B-components are given by:

\begin{equation}\label{eq:A}
    A = \frac{R_{PHE}I_{0}\tau_{FL}/\gamma}{H}
\end{equation}

\begin{equation}\label{eq:B}
    B = \frac{R_{AHE}I_{0}\tau_{DL}/\gamma}{H + H_{K}} + I_{0}R_{ANE}.
\end{equation}

\noindent Here, $\gamma$ is the gyromagnetic ratio, $H_{K}$ is the out-of-plane anisotropy field, and $R_{ANE}$ is the anomalous Nernst resistance. 
The SHH is fitted using Eq.(\ref{eq:SHH}) to extract the amplitude of the $\cos(\phi)$ and $\cos(2\phi)\cos(\phi)$ components, referred to as the A- and B-components, respectively.
As can be seen from Fig. \ref{fig:figure1}(e), our data is well described by Eq.(\ref{eq:SHH}).
Subsequently, the A- and B-components are determined for different magnetic fields, allowing us to extract the $\tau_{FL}$, $\tau_{DL}$ and anomalous Nernst resistance ($R_{ANE}$) using Eq.(\ref{eq:A}) and Eq.(\ref{eq:B}).
For the MoS\textsubscript{2}/Py/Al\textsubscript{2}O\textsubscript{3} devices, we corrected our data for a systematic 15 mT offset in the field.
The $R_{AHE} = 0.15$ $\Omega$ is obtained by performing a separate measurement where the first harmonic Hall voltage is measured while sweeping the magnetic field out-of-plane from approximately -1 T to 1 T.

Figure \ref{fig:figure2}(a) and (c) show the field-dependence of the A- and B-component of a MOCVD MoS\textsubscript{2}/Py/Al\textsubscript{2}O\textsubscript{3} device, respectively. 
As expected from Eq.(\ref{eq:A}), the A-component shows a linear dependence on the inverse magnetic field, which clearly indicates the presence of a field-like torque $\tau_{FL}$. 
Using Eq.(\ref{eq:A}) we find a $\tau_{FL}/\gamma$ ranging from
$\left(-2.5 \pm 0.2 \right) \times 10^{-6}$
T to
$\left(-7.2 \pm 0.3\right) \times 10^{-6}$
T, for an applied current density of 0.8$\times10^{10}$ A/m$^{2}$ to 5$\times10^{10}$ A/m$^{2}$.
This corresponds to a spin-torque conductivity $\sigma_{FL}$ of
$\left(-7.5 \pm 0.6 \right) \times10^{3} \frac{\hbar}{2e} (\Omega\cdot \mathrm{m})^{-1}$
to 
$\left(-3.5 \pm 0.2 \right) \times10^{3} \frac{\hbar}{2e} (\Omega\cdot \mathrm{m})^{-1}$,
respectively (see the Methods section for the spin-torque conductivity calculation).
This value is comparable to previous reports on chemical vapor deposition (CVD) grown MoS\textsubscript{2}/CoFeB devices from Shao et al. \cite{Shao2016e} and slightly lower compared to other semiconducting TMD/Py devices (e.g. WS\textsubscript{2}, WSe\textsubscript{2}) \cite{Shao2016e,Lv2018a, Hidding2021c, Novakov2021c}.
Larger field-like torques are reported in semimetallic TMD/Py bilayers, which is explained by the considerable Oersted torque arising from the current flowing through the conducting TMD \cite{MacNeill2017a, Guimaraes2018, Stiehl2019d}.
Due to the semiconducting character of the MoS\textsubscript{2} layer in our devices, no current is expected to flow through the MoS\textsubscript{2} and thus no Oersted torque is expected.

The field dependence of the B-component is depicted in Fig. \ref{fig:figure2}(c).
In the presence of a $\tau_{DL}$, a linear dependence is expected versus the inverse field (see Eq.(\ref{eq:B}), similar to the A-component).
At low magnetic fields, however, we observe large deviations from the linear dependence.
This can be explained by the fact that Eq.(\ref{eq:A}) and Eq.(\ref{eq:B}) are derived assuming that $H \gg H_{A}$, where $H_{A}$ is the in-plane uniaxial anisotropy field, which does not hold anymore at low magnetic fields \cite{Hayashi2014b}.
To get a more accurate estimate of the damping-like torque, we therefore neglect the three data point at lowest field when fitting the data to Eq.(\ref{eq:B}).
For the majority of our devices, we find a $\tau_{DL}/\gamma$ with a large error, ranging from
$\left(-4 \pm 33 \right) \times 10^{-6}$
T to
$\left(1.2 \pm 0.8 \right) \times 10^{-4}$
T, corresponding to a spin-torque conductivity $\sigma_{DL}$ ranging from
$\left(-4 \pm 32 \right) \times 10^{3} \frac{\hbar}{2e} (\Omega\cdot \mathrm{m})^{-1}$
to 
$\left(3 \pm 3 \right) \times 10^{5} \frac{\hbar}{2e} (\Omega\cdot \mathrm{m})^{-1}$.
Due to the large errors, we cannot conclude that these devices show a significant damping-like torque.
For one of the samples, however, we find a damping-like torque with a smaller error
$\tau_{DL}/\gamma = \left(10 \pm 2 \right) \times 10^{-5}$
T, corresponding to
$\sigma_{DL} = \left(1.2 \pm 0.3 \right) \times10^{5} \frac{\hbar}{2e} (\Omega\cdot \mathrm{m})^{-1}$.
This value is comparable to values reported in SOT devices made with Pt, W, and NiPS\textsubscript{3}, and are significantly higher compared to SOT devices using more conventional TMDs (e.g. WTe\textsubscript{2}, MoTe\textsubscript{2}, WSe\textsubscript{2}, etc.).
The presence of both a field-like and damping-like torque in MoS\textsubscript{2}/Py has been previously reported by Zhang et al. in ST-FMR measurements \cite{Zhang2016a}.
There, a torque ratio of $\tau_{FL}/\tau_{DL}=0.19\pm 0.01$ is reported, indicating a 5 times stronger damping-like torque.
Similarly, we find a stronger damping-like torque for this one device, showing a damping-like torque a factor of 20 stronger than the field-like torque.
On the other hand, Shao et al. report no damping-like torque in their SHH measurements on MoS\textsubscript{2}/CoFeB bilayers \cite{Shao2016e}, similar to our other devices.
These contrasting observations show that there is a significant device-to-device variation for the damping-like torque in these bilayers.

\subsection{Single-layer Py/Al\textsubscript{2}O\textsubscript{3} device}
To determine the contribution of possible self-torques in the Py-layer and accurately resolve the effect of the MOCVD grown MoS\textsubscript{2} layer on the SOTs, we compare the SOTs measurements from the MoS\textsubscript{2}/Py/Al\textsubscript{2}O\textsubscript{3} device to a single-layer Py/Al\textsubscript{2}O\textsubscript{3} reference device.
In Fig. \ref{fig:figure2}(b) and (d) the extracted A- and B-components for the single-layer Py/Al\textsubscript{2}O\textsubscript{3} device are plotted versus the inverse field. 
Surprisingly, even without the MoS\textsubscript{2} layer, we observe a clear linear dependence for the A-component  similar to the MoS\textsubscript{2}/Py/Al\textsubscript{2}O\textsubscript{3} devices, indicating the presence of a field-like self-torque.
Using Eq.(\ref{eq:A}) and Eq.(\ref{eq:spintorqueconductivity}), we find 
$\tau_{FL}/\gamma=\left(-8.6\pm 0.9 \right) \times 10^{-6}$ T 
and 
$\sigma_{FL}= \left(-2.8 \pm 0.3 \right) \times10^{3} \frac{\hbar}{2e} (\Omega\cdot \mathrm{m})^{-1}$.
The $\sigma_{FL}$ has the same sign and its magnitude is only $25\%$ lower compared to the MoS\textsubscript{2}/Py/Al\textsubscript{2}O\textsubscript{3} device, which indicates that the presence of the TMD layer does not significantly enhance the field-like SOT conductivity.

For the B-component, however, no large deviations at low fields are observed, as was the case with the MoS\textsubscript{2}/Py/Al\textsubscript{2}O\textsubscript{3} device.
Using Eq.(\ref{eq:A}) and Eq.(\ref{eq:spintorqueconductivity}), we find
$\tau_{DL}/\gamma = \left(-8 \pm 2 \right) \times 10^{-4}$ T 
and
$\sigma_{DL} = \left(-2.6 \pm 0.6\right) \times10^{5} \frac{\hbar}{2e} (\Omega\cdot \mathrm{m})^{-1}$
for the damping-like torque, which is larger and has an opposite sign compared to the MoS\textsubscript{2}/Py/Al\textsubscript{2}O\textsubscript{3} device that did show a significant damping-like torque.
For the MoS\textsubscript{2}/Py/Al\textsubscript{2}O\textsubscript{3} devices, the self-torque from the Py and the torque arising from the TMD/Py interaction might both be present.
As these torques have opposite sign, their competition reduces the net damping-like torque, which could explain the absence of a damping-like torque in the majority of our MoS\textsubscript{2}/Py/Al\textsubscript{2}O\textsubscript{3} devices.
This suggests that the addition of the TMD layer causes a suppression of the $\tau_{DL}$ rather than an increase, as was reported by Zhang et al. \cite{Zhang2016a} in CVD grown MoS\textsubscript{2}/Py devices.
Furthermore, differences in the interface quality, resulting in different contributions of the self-torque in Py and the torque arising from the TMD/FM interface could explain the contrasting SOT observations in MoS\textsubscript{2}/CoFeB and MoS\textsubscript{2}/Py  devices \cite{Shao2016e, Zhang2016a}.

Other reports on single-layer Py devices show both a damping-like and a field-like torque in ST-FMR measurements \cite{Seki2021a}.
Although a field-like torque was observed in all devices, only a damping-like torque was observed in devices where the structural inversion symmetry was broken.
In our devices, the structural inversion symmetry is broken as well, as the Py is evaporated on SiO\textsubscript{2} and capped with Al\textsubscript{2}O\textsubscript{3}, and thus possesses two different interfaces.
A difference in electron-scattering from these two interfaces could, in turn, lead to a self-torque.
Furthermore, Seki et al. report only damping-like torques for devices where the Py layer is sufficiently thin ($\le3$ nm), which is in contrast to our observation, as we observe a rather strong damping-like torque with a Py thickness of 6 nm \cite{Seki2021a}.
Also, Schippers et al. report measurements on a similar single-layer Py reference sample with a 6 nm Py thickness, capped with Al\textsubscript{2}O\textsubscript{3} \cite{Schippers2020c}.
At room temperature, they find a $\sigma_{FL}$ which is 3 times larger, and a $\sigma_{DL}$ which is one order of magnitude smaller.
For their samples, however, the layers are deposited using magnetron sputtering, while our samples employed electron beam evaporation, which could lead to different material and interface qualities, and different current distributions in the Py layer.

All these different torque strengths and directions observed for similar MoS\textsubscript{2}/FM bilayers and single-layer Py devices underline the large device-to-device variation, also observed in our devices.
Our observations show that the self-torque, originating solely from the FM layer, can have a significant contribution to the observed SOTs in TMD/FM bilayers.

\subsection{Effect of the Hall bar dimensions}
\begin{figure}[t]
	\includegraphics{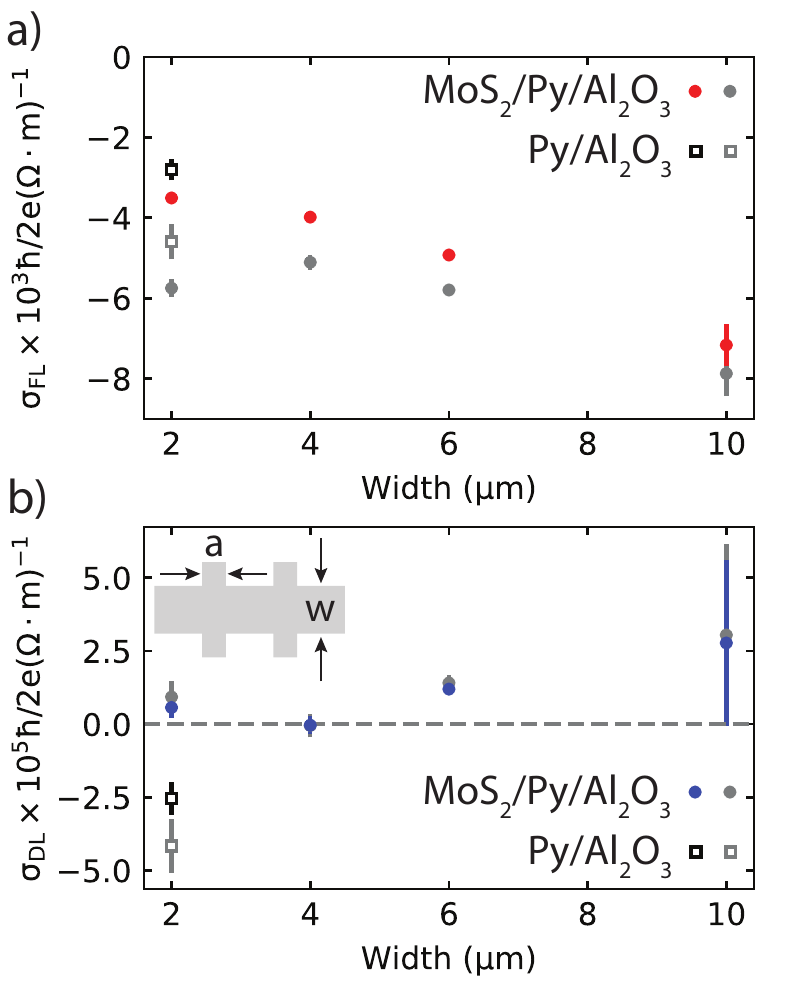}
	\caption{(a) The field-like and (b) damping-like spin-torque conductivity for the MoS\textsubscript{2}/Py/Al\textsubscript{2}O\textsubscript{3} devices (circles) with different channel width, and the single-layer Py/Al\textsubscript{2}O\textsubscript{3} device (squares). The colored points correspond to the raw spin-torque conductivity and the gray points to the corrected spin-torque conductivity's, according to ref\cite{Neumann2018a}. The spin-torque conductivity's of the single-layer Py/Al\textsubscript{2}O\textsubscript{3} device are indicated with the squares. The inset in (a) shows the channel width $w$ and width of the voltage legs ($a$)}
	\label{fig:figure3}
\end{figure}

Lastly, we study the effect of the Hall bar leg width/channel width ratio for the MoS\textsubscript{2}/Py/Al\textsubscript{2}O\textsubscript{3} devices by keeping the leg width constant at $a = $2 $\upmu$m, while varying the channel width ($w$) from 2 $\upmu$m to 10 $\upmu$m, plotted in Fig. \ref{fig:figure3}.
The extracted field-like (red circles) and damping-like spin-torque conductivities (blue circle) for all devices are plotted versus the channel width in Fig. \ref{fig:figure3}(a) and (b), respectively.
The $\sigma_{FL(DL)}$ from the single-layer Py/Al\textsubscript{2}O\textsubscript{3} device are included and depicted by the black unfilled squares for comparison.
We observe a clear, almost linear, dependence of $\sigma_{FL}$ on the channel width.
The $\sigma_{FL}$ for the 10 $\upmu$m device is a factor 2 larger than the 2 $\upmu$m device.
We stress that the larger error bar for the device with a channel width of 10 $\upmu$m is ascribed to a smaller current density compared to the other devices.
This observation is in line with recent work from Neumann et al. where the leg width/channel width ratio is shown to affect the estimation of the spin-Hall angle ($\theta_{SH}$) \cite{Neumann2018a}.
A significantly decreasing $\theta_{SH}$ is found when the leg width/channel width ratio becomes sufficiently big ($\geq1$), reporting a value of only $70\%$ at a leg width/channel width ratio of 1.
To correct our $\sigma_{FL(DL)}$, we incorporate a factor for each leg width/channel width ratio as reported by Neumann et al. \cite{Neumann2018a}, shown in Fig. \ref{fig:figure3} as gray circles.

After the correction, there is no clear monotonic decrease of $\sigma_{FL}$ with the leg width/channel width ratio.
However, still some device-to-device variation is found, which could be due to varying interface and material qualities.
For $\sigma_{DL}$, the reported values for both the MoS\textsubscript{2}/Py/Al\textsubscript{2}O\textsubscript{3} and the single-layer Py/Al\textsubscript{2}O\textsubscript{3} devices remain large with correspondingly large error bars as previously discussed.

\section{Conclusion}\label{sec:conclusion}
In conclusion, our results indicate that the self-torques, originating from the FM layer, can have significant contributions to the observed SOTs in TMD/FM bilayers.
We find a significant $\tau_{DL}$ in our Py/Al\textsubscript{2}O\textsubscript{3} reference device with an opposite sign compared to the $\tau_{DL}$ found in a MoS\textsubscript{2}/Py/Al\textsubscript{2}O\textsubscript{3} device.
These findings suggest that the self-torque from the FM layer can compete with torques originating from the presence of the TMD layer, reducing the torque strength in TMD/FM bilayers.
In addition, we observe a similar $\tau_{FL}$ in single-layer Py/Al\textsubscript{2}O\textsubscript{3} devices,
indicating that the MOCVD MoS\textsubscript{2} layer in our MoS\textsubscript{2}/Py/Al\textsubscript{2}O\textsubscript{3} is of minor importance for the generation of a field-like SOTs.
Both findings suggest that the FM layer, rather than the TMD, might play a dominant role in the generation of the observed SOTs in TMD/FM bilayers.
Previous contrasting SOT observations in similar TMD/FM bilayers could thus be ascribed to differences in the FM layer (e.g., different fabrication techniques, capping layers, interface quality, etc.) rather than different interactions between the TMD and the FM.
We therefore recommend that the self-torque in a single-layer FM reference samples are also characterized in future studies in order to accurately determine the effect of the TMD layer on the observed SOTs, and that the Hall bar dimensions should be clearly reported.
In addition, in HM/FM bilayers, the self-torque in the FM layers could counteract the torque generated by the HM layer, resulting in a reduced net torque.
Taking advantage of the self-torques in the FM layer, making them work in conjunction with other SOTs instead, could lead to an increase in the SOT efficiency \cite{Aoki2022}.
Tailoring the FM interfaces to change the self-torque direction and strength, in conjunction with searching for different materials as SOT sources, could be a promising route towards an increase in SOT efficiency.
These results pave the way for a more accurate disentanglement of all microscopic mechanisms at play, increasing our understanding of the origins underlying the SOTs, which is essential for more energy efficient magnetic memory devices.

\section{Methods}\label{sec:methods}
\subsection{Device fabrication}
The MoS\textsubscript{2} layer was grown using metal-organic chemical vapor deposition (MOCVD) on a SiO\textsubscript{2}/Si substrate as described in ref\cite{Kang2015}.
The  MoS\textsubscript{2} is characterized with a PL microscope using a BrightLine long-pass filter set to check the homogeneity of the monolayer coverage on the SiO\textsubscript{2}/Si substrate (see Fig. \ref{fig:figure1}).

Next, a prepared PMMA mask with exposed Hall bars of different widths is deposited on top of the MoS\textsubscript{2} covered substrate, which ensures a pristine interface between the permalloy and MoS\textsubscript{2} with no polymer contamination.
Using electron beam evaporation, 6 nm of permalloy and a capping layer and hard mask of 17 nm of Al\textsubscript{2}O\textsubscript{3} are deposited.
Subsequently, the contacts are defined using standard e-beam lithography techniques.
Then, first an Al\textsubscript{2}O\textsubscript{3} wet etch with tetramethylammonium is performed for 45 seconds at 40 $^\circ$C, after which in-situ Ar-milling is performed prior to the evaporation of the Ti/Au (5/55 nm) contacts.
Finally, the remaining MoS\textsubscript{2} layer is removed using reactive ion etching (30 W RF, 5W ICP).

\subsection{Electrical measurements}
The harmonic Hall measurements, illustrated in Fig. \ref{fig:figure1}a and Fig. \ref{fig:figure1}c, were performed at room temperature using a standard lock-in technique with low frequency (77.77 Hz) AC-currents ($I_{0}$), ranging from $500$ $\upmu$A to $700$ $\upmu$A \cite{Garello2013d, Hayashi2014b, Nguyen2021a}.
Subsequently, the first ($V_{xy}^{\omega}$) and second ($V_{xy}^{2\omega}$) harmonic Hall voltage were measured while an applied magnetic field ($H$), ranging from 10 mT to 300 mT, was rotated in-plane, making an angle $\phi$ with respect to the current (Fig. \ref{fig:figure1}(c)).

To better compare the SOTs in our devices to previous reports on SOTs in TMD/FM bilayers, we express the SOT in terms of spin-torque conductivity; the common figure-of-merit in literature due to its independence on geometric factors \cite{Nguyen2016a, Stiehl2019d}. The spin-torque conductivity is defined as the total angular momentum absorbed by the ferromagnet per second, per unit interface area, per applied electric field, in units of $\frac{\hbar}{2e}$, and is calculated according to:

\begin{equation} \label{eq:spintorqueconductivity}
\begin{split}
\sigma_{FL(DL)} & = \frac{2e}{\hbar} M_{s}t_{py} (lw) \frac{\tau_{FL(DL)}/\gamma}{(lw)E}\\
 & = \frac{2e}{\hbar} M_{s}t_{Py}w\frac{\tau_{FL(DL)}/\gamma}{R_{sq}I_{0}},
\end{split}
\end{equation}

\noindent where $M_{s}$ is the saturation magnetization, $\hbar$ is the reduced Planks constant, $e$ is the electron charge, $E$ is the electric field, $R_{sq}$ is the square resistance, $I_{0}$ is the applied current, and $l$, $w$, and $t_{Py}$ are the length, width and Py thickness, respectively.

\subsection{Anomalous Hall measurement}
\begin{figure}[h!]
	\includegraphics[width=0.4\textwidth]{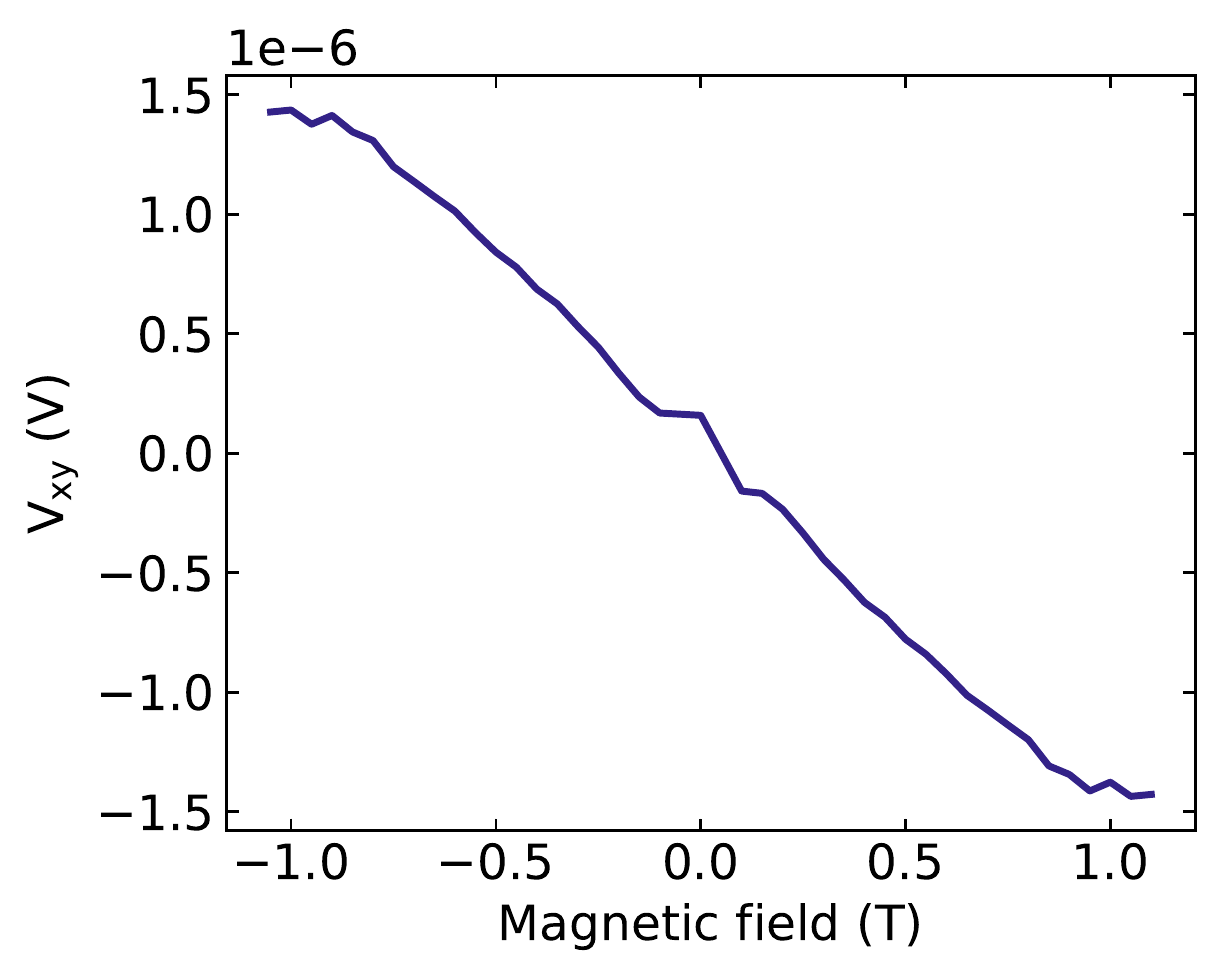}
	\caption{The antisymmetrized Hall voltage for the anomalous Hall measurement used to extract the anomalous Hall resistance, $R_{AHE}$, and the saturation magnetization, $M_{s}$, needed to determine $\tau_{DL}$ from Eq.(\ref{eq:B}). The data presented here is antisymmetrized to reduce any errors due to sample misalignment.}
	\label{fig:figure4}
\end{figure}

To determine the anomalous Hall resistance, $R_{AHE}$, and the saturation magnetization, $M_{s}$, needed for determining the damping-like torque $\tau_{DL}$ using Eq.\ref{eq:B}, we performed anomalous Hall measurements (Fig. \ref{fig:figure4}).
Using a standard lock-in technique with a low frequency (17.77 Hz) current of 1 \textmu A, the Hall voltage is measured while an out-of-plane magnetic field is sweeped from $\sim$-1 T to 1 T, as shown in Fig. \ref{fig:figure4}.
To reduce errors from any misalignment, the Hall voltage is antisymmetrized.

\section{Acknowledgements}
We would like to acknowledge Prof. M. A. Loi and E. K. Tekelenburg for their help with the Raman measurements and thank J. G. Holstein, H. Adema, H. de Vries, A. Joshua, and F. H. van der Velde for their technical support.
Sample fabrication was performed using NanoLabNL facilities.
This work was supported by the Dutch Research Council (NWO—STU.019.014), the Zernike Institute for Advanced Materials, and innovation program under grant agreement No. 881603 (Graphene Flagship).

\section{Author contributions}
J.H. and K.M. fabricated the samples, and performed both the electrical and optical measurements under supervision of M.H.D.G..
F.M. and C.L. grew the MOCVD MoS\textsubscript{2} layer under supervision of J.P.
J.H. performed the data analysis under supervision of M.H.D.G.
J.H. and M.H.D.G. wrote the paper with comments from all authors.


\begin{thebibliography}{50}%
	\makeatletter
	\providecommand \@ifxundefined [1]{%
		\@ifx{#1\undefined}
	}%
	\providecommand \@ifnum [1]{%
		\ifnum #1\expandafter \@firstoftwo
		\else \expandafter \@secondoftwo
		\fi
	}%
	\providecommand \@ifx [1]{%
		\ifx #1\expandafter \@firstoftwo
		\else \expandafter \@secondoftwo
		\fi
	}%
	\providecommand \natexlab [1]{#1}%
	\providecommand \enquote  [1]{``#1''}%
	\providecommand \bibnamefont  [1]{#1}%
	\providecommand \bibfnamefont [1]{#1}%
	\providecommand \citenamefont [1]{#1}%
	\providecommand \href@noop [0]{\@secondoftwo}%
	\providecommand \href [0]{\begingroup \@sanitize@url \@href}%
	\providecommand \@href[1]{\@@startlink{#1}\@@href}%
	\providecommand \@@href[1]{\endgroup#1\@@endlink}%
	\providecommand \@sanitize@url [0]{\catcode `\\12\catcode `\$12\catcode
		`\&12\catcode `\#12\catcode `\^12\catcode `\_12\catcode `\%12\relax}%
	\providecommand \@@startlink[1]{}%
	\providecommand \@@endlink[0]{}%
	\providecommand \url  [0]{\begingroup\@sanitize@url \@url }%
	\providecommand \@url [1]{\endgroup\@href {#1}{\urlprefix }}%
	\providecommand \urlprefix  [0]{URL }%
	\providecommand \Eprint [0]{\href }%
	\providecommand \doibase [0]{http://dx.doi.org/}%
	\providecommand \selectlanguage [0]{\@gobble}%
	\providecommand \bibinfo  [0]{\@secondoftwo}%
	\providecommand \bibfield  [0]{\@secondoftwo}%
	\providecommand \translation [1]{[#1]}%
	\providecommand \BibitemOpen [0]{}%
	\providecommand \bibitemStop [0]{}%
	\providecommand \bibitemNoStop [0]{.\EOS\space}%
	\providecommand \EOS [0]{\spacefactor3000\relax}%
	\providecommand \BibitemShut  [1]{\csname bibitem#1\endcsname}%
	\let\auto@bib@innerbib\@empty
	\bibitem [{\citenamefont {Akyol}\ \emph {et~al.}(2016)\citenamefont {Akyol},
		\citenamefont {Jiang}, \citenamefont {Yu}, \citenamefont {Fan}, \citenamefont
		{Gunes}, \citenamefont {Ekicibil}, \citenamefont {{Khalili Amiri}},\ and\
		\citenamefont {Wang}}]{Akyol2016}%
	\BibitemOpen
	\bibfield  {author} {\bibinfo {author} {\bibnamefont {Akyol}, \bibfnamefont
			{M.}}, \bibinfo {author} {\bibnamefont {Jiang}, \bibfnamefont {W.}}, \bibinfo
		{author} {\bibnamefont {Yu}, \bibfnamefont {G.}}, \bibinfo {author}
		{\bibnamefont {Fan}, \bibfnamefont {Y.}}, \bibinfo {author} {\bibnamefont
			{Gunes}, \bibfnamefont {M.}}, \bibinfo {author} {\bibnamefont {Ekicibil},
			\bibfnamefont {A.}}, \bibinfo {author} {\bibnamefont {{Khalili Amiri}},
			\bibfnamefont {P.}}, \ and\ \bibinfo {author} {\bibnamefont {Wang},
			\bibfnamefont {K.~L.}},\ }\href {\doibase 10.1063/1.4958295} {\bibfield
		{journal} {\bibinfo  {journal} {Applied Physics Letters}\ }\textbf {\bibinfo
			{volume} {109}},\ \bibinfo {pages} {022403} (\bibinfo {year}
		{2016})}\BibitemShut {NoStop}%
	\bibitem [{\citenamefont {Amin}, \citenamefont {Haney},\ and\ \citenamefont
		{Stiles}(2020)}]{Amin2020}%
	\BibitemOpen
	\bibfield  {author} {\bibinfo {author} {\bibnamefont {Amin}, \bibfnamefont
			{V.~P.}}, \bibinfo {author} {\bibnamefont {Haney}, \bibfnamefont {P.~M.}}, \
		and\ \bibinfo {author} {\bibnamefont {Stiles}, \bibfnamefont {M.~D.}},\
	}\href {\doibase 10.1063/5.0024019} {\bibfield  {journal} {\bibinfo
			{journal} {Journal of Applied Physics}\ }\textbf {\bibinfo {volume} {128}}
		(\bibinfo {year} {2020}),\ 10.1063/5.0024019},\ \Eprint
	{http://arxiv.org/abs/2008.01182} {arXiv:2008.01182} \BibitemShut {NoStop}%
	\bibitem [{\citenamefont {Aoki}\ \emph {et~al.}(2022)\citenamefont {Aoki},
		\citenamefont {Shigematsu}, \citenamefont {Ohshima}, \citenamefont {Shinjo},
		\citenamefont {Shiraishi},\ and\ \citenamefont {Ando}}]{Aoki2022}%
	\BibitemOpen
	\bibfield  {author} {\bibinfo {author} {\bibnamefont {Aoki}, \bibfnamefont
			{M.}}, \bibinfo {author} {\bibnamefont {Shigematsu}, \bibfnamefont {E.}},
		\bibinfo {author} {\bibnamefont {Ohshima}, \bibfnamefont {R.}}, \bibinfo
		{author} {\bibnamefont {Shinjo}, \bibfnamefont {T.}}, \bibinfo {author}
		{\bibnamefont {Shiraishi}, \bibfnamefont {M.}}, \ and\ \bibinfo {author}
		{\bibnamefont {Ando}, \bibfnamefont {Y.}},\ }\href {\doibase
		10.1103/PhysRevB.106.174418} {\bibfield  {journal} {\bibinfo  {journal}
			{Physical Review B}\ }\textbf {\bibinfo {volume} {106}},\ \bibinfo {pages}
		{174418} (\bibinfo {year} {2022})}\BibitemShut {NoStop}%
	\bibitem [{\citenamefont {Avci}\ \emph {et~al.}(2014)\citenamefont {Avci},
		\citenamefont {Garello}, \citenamefont {Gabureac}, \citenamefont {Ghosh},
		\citenamefont {Fuhrer}, \citenamefont {Alvarado},\ and\ \citenamefont
		{Gambardella}}]{Avci2014b}%
	\BibitemOpen
	\bibfield  {author} {\bibinfo {author} {\bibnamefont {Avci}, \bibfnamefont
			{C.~O.}}, \bibinfo {author} {\bibnamefont {Garello}, \bibfnamefont {K.}},
		\bibinfo {author} {\bibnamefont {Gabureac}, \bibfnamefont {M.}}, \bibinfo
		{author} {\bibnamefont {Ghosh}, \bibfnamefont {A.}}, \bibinfo {author}
		{\bibnamefont {Fuhrer}, \bibfnamefont {A.}}, \bibinfo {author} {\bibnamefont
			{Alvarado}, \bibfnamefont {S.~F.}}, \ and\ \bibinfo {author} {\bibnamefont
			{Gambardella}, \bibfnamefont {P.}},\ }\href {\doibase
		10.1103/PhysRevB.90.224427} {\bibfield  {journal} {\bibinfo  {journal}
			{Physical Review B}\ }\textbf {\bibinfo {volume} {90}},\ \bibinfo {pages}
		{224427} (\bibinfo {year} {2014})},\ \Eprint {http://arxiv.org/abs/1412.0865}
	{arXiv:1412.0865} \BibitemShut {NoStop}%
	\bibitem [{\citenamefont {Demasius}\ \emph {et~al.}(2016)\citenamefont
		{Demasius}, \citenamefont {Phung}, \citenamefont {Zhang}, \citenamefont
		{Hughes}, \citenamefont {Yang}, \citenamefont {Kellock}, \citenamefont {Han},
		\citenamefont {Pushp},\ and\ \citenamefont {Parkin}}]{Demasius2016}%
	\BibitemOpen
	\bibfield  {author} {\bibinfo {author} {\bibnamefont {Demasius},
			\bibfnamefont {K.-u.}}, \bibinfo {author} {\bibnamefont {Phung},
			\bibfnamefont {T.}}, \bibinfo {author} {\bibnamefont {Zhang}, \bibfnamefont
			{W.}}, \bibinfo {author} {\bibnamefont {Hughes}, \bibfnamefont {B.~P.}},
		\bibinfo {author} {\bibnamefont {Yang}, \bibfnamefont {S.-H.}}, \bibinfo
		{author} {\bibnamefont {Kellock}, \bibfnamefont {A.}}, \bibinfo {author}
		{\bibnamefont {Han}, \bibfnamefont {W.}}, \bibinfo {author} {\bibnamefont
			{Pushp}, \bibfnamefont {A.}}, \ and\ \bibinfo {author} {\bibnamefont
			{Parkin}, \bibfnamefont {S.~S.~P.}},\ }\href {\doibase 10.1038/ncomms10644}
	{\bibfield  {journal} {\bibinfo  {journal} {Nature Communications}\ }\textbf
		{\bibinfo {volume} {7}},\ \bibinfo {pages} {10644} (\bibinfo {year}
		{2016})}\BibitemShut {NoStop}%
	\bibitem [{\citenamefont {Dieny}\ \emph {et~al.}(2020)\citenamefont {Dieny},
		\citenamefont {Prejbeanu}, \citenamefont {Garello}, \citenamefont
		{Gambardella}, \citenamefont {Freitas}, \citenamefont {Lehndorff},
		\citenamefont {Raberg}, \citenamefont {Ebels}, \citenamefont {Demokritov},
		\citenamefont {Akerman}, \citenamefont {Deac}, \citenamefont {Pirro},
		\citenamefont {Adelmann}, \citenamefont {Anane}, \citenamefont {Chumak},
		\citenamefont {Hirohata}, \citenamefont {Mangin}, \citenamefont {Valenzuela},
		\citenamefont {Onbaşlı}, \citenamefont {D'Aquino}, \citenamefont {Prenat},
		\citenamefont {Finocchio}, \citenamefont {Lopez-Diaz}, \citenamefont
		{Chantrell}, \citenamefont {Chubykalo-Fesenko},\ and\ \citenamefont
		{Bortolotti}}]{Dieny2020b}%
	\BibitemOpen
	\bibfield  {author} {\bibinfo {author} {\bibnamefont {Dieny}, \bibfnamefont
			{B.}}, \bibinfo {author} {\bibnamefont {Prejbeanu}, \bibfnamefont {I.~L.}},
		\bibinfo {author} {\bibnamefont {Garello}, \bibfnamefont {K.}}, \bibinfo
		{author} {\bibnamefont {Gambardella}, \bibfnamefont {P.}}, \bibinfo {author}
		{\bibnamefont {Freitas}, \bibfnamefont {P.}}, \bibinfo {author} {\bibnamefont
			{Lehndorff}, \bibfnamefont {R.}}, \bibinfo {author} {\bibnamefont {Raberg},
			\bibfnamefont {W.}}, \bibinfo {author} {\bibnamefont {Ebels}, \bibfnamefont
			{U.}}, \bibinfo {author} {\bibnamefont {Demokritov}, \bibfnamefont {S.~O.}},
		\bibinfo {author} {\bibnamefont {Akerman}, \bibfnamefont {J.}}, \bibinfo
		{author} {\bibnamefont {Deac}, \bibfnamefont {A.}}, \bibinfo {author}
		{\bibnamefont {Pirro}, \bibfnamefont {P.}}, \bibinfo {author} {\bibnamefont
			{Adelmann}, \bibfnamefont {C.}}, \bibinfo {author} {\bibnamefont {Anane},
			\bibfnamefont {A.}}, \bibinfo {author} {\bibnamefont {Chumak}, \bibfnamefont
			{A.~V.}}, \bibinfo {author} {\bibnamefont {Hirohata}, \bibfnamefont {A.}},
		\bibinfo {author} {\bibnamefont {Mangin}, \bibfnamefont {S.}}, \bibinfo
		{author} {\bibnamefont {Valenzuela}, \bibfnamefont {S.~O.}}, \bibinfo
		{author} {\bibnamefont {Onbaşlı}, \bibfnamefont {M.~C.}}, \bibinfo {author}
		{\bibnamefont {D'Aquino}, \bibfnamefont {M.}}, \bibinfo {author}
		{\bibnamefont {Prenat}, \bibfnamefont {G.}}, \bibinfo {author} {\bibnamefont
			{Finocchio}, \bibfnamefont {G.}}, \bibinfo {author} {\bibnamefont
			{Lopez-Diaz}, \bibfnamefont {L.}}, \bibinfo {author} {\bibnamefont
			{Chantrell}, \bibfnamefont {R.}}, \bibinfo {author} {\bibnamefont
			{Chubykalo-Fesenko}, \bibfnamefont {O.}}, \ and\ \bibinfo {author}
		{\bibnamefont {Bortolotti}, \bibfnamefont {P.}},\ }\href {\doibase
		10.1038/s41928-020-0461-5} {\bibfield  {journal} {\bibinfo  {journal} {Nature
				Electronics}\ }\textbf {\bibinfo {volume} {3}},\ \bibinfo {pages} {446}
		(\bibinfo {year} {2020})}\BibitemShut {NoStop}%
	\bibitem [{\citenamefont {Fan}\ \emph {et~al.}(2014)\citenamefont {Fan},
		\citenamefont {Celik}, \citenamefont {Wu}, \citenamefont {Ni}, \citenamefont
		{Lee}, \citenamefont {Lorenz},\ and\ \citenamefont {Xiao}}]{Fan2014}%
	\BibitemOpen
	\bibfield  {author} {\bibinfo {author} {\bibnamefont {Fan}, \bibfnamefont
			{X.}}, \bibinfo {author} {\bibnamefont {Celik}, \bibfnamefont {H.}}, \bibinfo
		{author} {\bibnamefont {Wu}, \bibfnamefont {J.}}, \bibinfo {author}
		{\bibnamefont {Ni}, \bibfnamefont {C.}}, \bibinfo {author} {\bibnamefont
			{Lee}, \bibfnamefont {K.-J.}}, \bibinfo {author} {\bibnamefont {Lorenz},
			\bibfnamefont {V.~O.}}, \ and\ \bibinfo {author} {\bibnamefont {Xiao},
			\bibfnamefont {J.~Q.}},\ }\href {\doibase 10.1038/ncomms4042} {\bibfield
		{journal} {\bibinfo  {journal} {Nature Communications}\ }\textbf {\bibinfo
			{volume} {5}},\ \bibinfo {pages} {3042} (\bibinfo {year} {2014})}\BibitemShut
	{NoStop}%
	\bibitem [{\citenamefont {Garello}\ \emph {et~al.}(2013)\citenamefont
		{Garello}, \citenamefont {Miron}, \citenamefont {Avci}, \citenamefont
		{Freimuth}, \citenamefont {Mokrousov}, \citenamefont {Bl{\"{u}}gel},
		\citenamefont {Auffret}, \citenamefont {Boulle}, \citenamefont {Gaudin},\
		and\ \citenamefont {Gambardella}}]{Garello2013d}%
	\BibitemOpen
	\bibfield  {author} {\bibinfo {author} {\bibnamefont {Garello}, \bibfnamefont
			{K.}}, \bibinfo {author} {\bibnamefont {Miron}, \bibfnamefont {I.~M.}},
		\bibinfo {author} {\bibnamefont {Avci}, \bibfnamefont {C.~O.}}, \bibinfo
		{author} {\bibnamefont {Freimuth}, \bibfnamefont {F.}}, \bibinfo {author}
		{\bibnamefont {Mokrousov}, \bibfnamefont {Y.}}, \bibinfo {author}
		{\bibnamefont {Bl{\"{u}}gel}, \bibfnamefont {S.}}, \bibinfo {author}
		{\bibnamefont {Auffret}, \bibfnamefont {S.}}, \bibinfo {author} {\bibnamefont
			{Boulle}, \bibfnamefont {O.}}, \bibinfo {author} {\bibnamefont {Gaudin},
			\bibfnamefont {G.}}, \ and\ \bibinfo {author} {\bibnamefont {Gambardella},
			\bibfnamefont {P.}},\ }\href {\doibase 10.1038/nnano.2013.145} {\bibfield
		{journal} {\bibinfo  {journal} {Nature Nanotechnology}\ }\textbf {\bibinfo
			{volume} {8}},\ \bibinfo {pages} {587} (\bibinfo {year} {2013})}\BibitemShut
	{NoStop}%
	\bibitem [{\citenamefont {Ghosh}\ \emph {et~al.}(2017)\citenamefont {Ghosh},
		\citenamefont {Garello}, \citenamefont {Avci}, \citenamefont {Gabureac},\
		and\ \citenamefont {Gambardella}}]{Ghosh2017b}%
	\BibitemOpen
	\bibfield  {author} {\bibinfo {author} {\bibnamefont {Ghosh}, \bibfnamefont
			{A.}}, \bibinfo {author} {\bibnamefont {Garello}, \bibfnamefont {K.}},
		\bibinfo {author} {\bibnamefont {Avci}, \bibfnamefont {C.~O.}}, \bibinfo
		{author} {\bibnamefont {Gabureac}, \bibfnamefont {M.}}, \ and\ \bibinfo
		{author} {\bibnamefont {Gambardella}, \bibfnamefont {P.}},\ }\href {\doibase
		10.1103/PhysRevApplied.7.014004} {\bibfield  {journal} {\bibinfo  {journal}
			{Physical Review Applied}\ }\textbf {\bibinfo {volume} {7}},\ \bibinfo
		{pages} {014004} (\bibinfo {year} {2017})},\ \Eprint
	{http://arxiv.org/abs/1701.01843} {arXiv:1701.01843} \BibitemShut {NoStop}%
	\bibitem [{\citenamefont {Guimar{\~{a}}es}\ \emph {et~al.}(2018)\citenamefont
		{Guimar{\~{a}}es}, \citenamefont {Stiehl}, \citenamefont {MacNeill},
		\citenamefont {Reynolds},\ and\ \citenamefont {Ralph}}]{Guimaraes2018}%
	\BibitemOpen
	\bibfield  {author} {\bibinfo {author} {\bibnamefont {Guimar{\~{a}}es},
			\bibfnamefont {M.~H.~D.}}, \bibinfo {author} {\bibnamefont {Stiehl},
			\bibfnamefont {G.~M.}}, \bibinfo {author} {\bibnamefont {MacNeill},
			\bibfnamefont {D.}}, \bibinfo {author} {\bibnamefont {Reynolds},
			\bibfnamefont {N.~D.}}, \ and\ \bibinfo {author} {\bibnamefont {Ralph},
			\bibfnamefont {D.~C.}},\ }\href {\doibase 10.1021/acs.nanolett.7b04993}
	{\bibfield  {journal} {\bibinfo  {journal} {Nano Letters}\ }\textbf {\bibinfo
			{volume} {18}},\ \bibinfo {pages} {1311} (\bibinfo {year} {2018})},\ \Eprint
	{http://arxiv.org/abs/1801.07281} {arXiv:1801.07281} \BibitemShut {NoStop}%
	\bibitem [{\citenamefont {Hayashi}\ \emph {et~al.}(2014)\citenamefont
		{Hayashi}, \citenamefont {Kim}, \citenamefont {Yamanouchi},\ and\
		\citenamefont {Ohno}}]{Hayashi2014b}%
	\BibitemOpen
	\bibfield  {author} {\bibinfo {author} {\bibnamefont {Hayashi}, \bibfnamefont
			{M.}}, \bibinfo {author} {\bibnamefont {Kim}, \bibfnamefont {J.}}, \bibinfo
		{author} {\bibnamefont {Yamanouchi}, \bibfnamefont {M.}}, \ and\ \bibinfo
		{author} {\bibnamefont {Ohno}, \bibfnamefont {H.}},\ }\href {\doibase
		10.1103/PhysRevB.89.144425} {\bibfield  {journal} {\bibinfo  {journal}
			{Physical Review B}\ }\textbf {\bibinfo {volume} {89}},\ \bibinfo {pages}
		{144425} (\bibinfo {year} {2014})}\BibitemShut {NoStop}%
	\bibitem [{\citenamefont {Hidding}\ and\ \citenamefont
		{Guimar{\~{a}}es}(2020)}]{Hidding2020b}%
	\BibitemOpen
	\bibfield  {author} {\bibinfo {author} {\bibnamefont {Hidding}, \bibfnamefont
			{J.}}\ and\ \bibinfo {author} {\bibnamefont {Guimar{\~{a}}es}, \bibfnamefont
			{M.~H.}},\ }\href {\doibase 10.3389/fmats.2020.594771} {\bibfield  {journal}
		{\bibinfo  {journal} {Frontiers in Materials}\ }\textbf {\bibinfo {volume}
			{7}} (\bibinfo {year} {2020}),\ 10.3389/fmats.2020.594771}\BibitemShut
	{NoStop}%
	\bibitem [{\citenamefont {Hidding}\ \emph {et~al.}(2021)\citenamefont
		{Hidding}, \citenamefont {Tirion}, \citenamefont {Momand}, \citenamefont
		{Kaverzin}, \citenamefont {Mostovoy}, \citenamefont {van Wees}, \citenamefont
		{Kooi},\ and\ \citenamefont {Guimar{\~{a}}es}}]{Hidding2021c}%
	\BibitemOpen
	\bibfield  {author} {\bibinfo {author} {\bibnamefont {Hidding}, \bibfnamefont
			{J.}}, \bibinfo {author} {\bibnamefont {Tirion}, \bibfnamefont {S.~H.}},
		\bibinfo {author} {\bibnamefont {Momand}, \bibfnamefont {J.}}, \bibinfo
		{author} {\bibnamefont {Kaverzin}, \bibfnamefont {A.}}, \bibinfo {author}
		{\bibnamefont {Mostovoy}, \bibfnamefont {M.}}, \bibinfo {author}
		{\bibnamefont {van Wees}, \bibfnamefont {B.~J.}}, \bibinfo {author}
		{\bibnamefont {Kooi}, \bibfnamefont {B.~J.}}, \ and\ \bibinfo {author}
		{\bibnamefont {Guimar{\~{a}}es}, \bibfnamefont {M.~H.}},\ }\href {\doibase
		10.1088/2515-7639/ac24ee} {\bibfield  {journal} {\bibinfo  {journal} {JPhys
				Materials}\ }\textbf {\bibinfo {volume} {4}} (\bibinfo {year} {2021}),\
		10.1088/2515-7639/ac24ee}\BibitemShut {NoStop}%
	\bibitem [{\citenamefont {Husain}\ \emph {et~al.}(2020)\citenamefont {Husain},
		\citenamefont {Gupta}, \citenamefont {Kumar}, \citenamefont {Kumar},
		\citenamefont {Behera}, \citenamefont {Brucas}, \citenamefont {Chaudhary},\
		and\ \citenamefont {Svedlindh}}]{Husain2020d}%
	\BibitemOpen
	\bibfield  {author} {\bibinfo {author} {\bibnamefont {Husain}, \bibfnamefont
			{S.}}, \bibinfo {author} {\bibnamefont {Gupta}, \bibfnamefont {R.}}, \bibinfo
		{author} {\bibnamefont {Kumar}, \bibfnamefont {A.}}, \bibinfo {author}
		{\bibnamefont {Kumar}, \bibfnamefont {P.}}, \bibinfo {author} {\bibnamefont
			{Behera}, \bibfnamefont {N.}}, \bibinfo {author} {\bibnamefont {Brucas},
			\bibfnamefont {R.}}, \bibinfo {author} {\bibnamefont {Chaudhary},
			\bibfnamefont {S.}}, \ and\ \bibinfo {author} {\bibnamefont {Svedlindh},
			\bibfnamefont {P.}},\ }\href {\doibase 10.1063/5.0025318} {\bibfield
		{journal} {\bibinfo  {journal} {Applied Physics Reviews}\ }\textbf {\bibinfo
			{volume} {7}} (\bibinfo {year} {2020}),\ 10.1063/5.0025318}\BibitemShut
	{NoStop}%
	\bibitem [{\citenamefont {Jamali}\ \emph {et~al.}(2013)\citenamefont {Jamali},
		\citenamefont {Narayanapillai}, \citenamefont {Qiu}, \citenamefont {Loong},
		\citenamefont {Manchon},\ and\ \citenamefont {Yang}}]{Jamali2013}%
	\BibitemOpen
	\bibfield  {author} {\bibinfo {author} {\bibnamefont {Jamali}, \bibfnamefont
			{M.}}, \bibinfo {author} {\bibnamefont {Narayanapillai}, \bibfnamefont {K.}},
		\bibinfo {author} {\bibnamefont {Qiu}, \bibfnamefont {X.}}, \bibinfo {author}
		{\bibnamefont {Loong}, \bibfnamefont {L.~M.}}, \bibinfo {author}
		{\bibnamefont {Manchon}, \bibfnamefont {A.}}, \ and\ \bibinfo {author}
		{\bibnamefont {Yang}, \bibfnamefont {H.}},\ }\href {\doibase
		10.1103/PhysRevLett.111.246602} {\bibfield  {journal} {\bibinfo  {journal}
			{Physical Review Letters}\ }\textbf {\bibinfo {volume} {111}},\ \bibinfo
		{pages} {246602} (\bibinfo {year} {2013})}\BibitemShut {NoStop}%
	\bibitem [{\citenamefont {Kang}\ \emph {et~al.}(2015)\citenamefont {Kang},
		\citenamefont {Xie}, \citenamefont {Huang}, \citenamefont {Han},
		\citenamefont {Huang}, \citenamefont {Mak}, \citenamefont {Kim},\ and\
		\citenamefont {Muller}}]{Kang2015}%
	\BibitemOpen
	\bibfield  {author} {\bibinfo {author} {\bibnamefont {Kang}, \bibfnamefont
			{K.}}, \bibinfo {author} {\bibnamefont {Xie}, \bibfnamefont {S.}}, \bibinfo
		{author} {\bibnamefont {Huang}, \bibfnamefont {L.}}, \bibinfo {author}
		{\bibnamefont {Han}, \bibfnamefont {Y.}}, \bibinfo {author} {\bibnamefont
			{Huang}, \bibfnamefont {P.~Y.}}, \bibinfo {author} {\bibnamefont {Mak},
			\bibfnamefont {K.~F.}}, \bibinfo {author} {\bibnamefont {Kim}, \bibfnamefont
			{C.-j.}}, \ and\ \bibinfo {author} {\bibnamefont {Muller}, \bibfnamefont
			{D.}},\ }\href {\doibase 10.1038/nature14417} {\bibfield  {journal} {\bibinfo
			{journal} {Nature}\ }\textbf {\bibinfo {volume} {520}},\ \bibinfo {pages}
		{656} (\bibinfo {year} {2015})}\BibitemShut {NoStop}%
	\bibitem [{\citenamefont {Kao}\ \emph {et~al.}(2022)\citenamefont {Kao},
		\citenamefont {Muzzio}, \citenamefont {Zhang}, \citenamefont {Zhu},
		\citenamefont {Gobbo}, \citenamefont {Yuan}, \citenamefont {Weber},
		\citenamefont {Rao}, \citenamefont {Li}, \citenamefont {Edgar}, \citenamefont
		{Goldberger}, \citenamefont {Yan}, \citenamefont {Mandrus}, \citenamefont
		{Hwang}, \citenamefont {Cheng}, \citenamefont {Katoch},\ and\ \citenamefont
		{Singh}}]{Kao2022}%
	\BibitemOpen
	\bibfield  {author} {\bibinfo {author} {\bibnamefont {Kao}, \bibfnamefont
			{I.~H.}}, \bibinfo {author} {\bibnamefont {Muzzio}, \bibfnamefont {R.}},
		\bibinfo {author} {\bibnamefont {Zhang}, \bibfnamefont {H.}}, \bibinfo
		{author} {\bibnamefont {Zhu}, \bibfnamefont {M.}}, \bibinfo {author}
		{\bibnamefont {Gobbo}, \bibfnamefont {J.}}, \bibinfo {author} {\bibnamefont
			{Yuan}, \bibfnamefont {S.}}, \bibinfo {author} {\bibnamefont {Weber},
			\bibfnamefont {D.}}, \bibinfo {author} {\bibnamefont {Rao}, \bibfnamefont
			{R.}}, \bibinfo {author} {\bibnamefont {Li}, \bibfnamefont {J.}}, \bibinfo
		{author} {\bibnamefont {Edgar}, \bibfnamefont {J.~H.}}, \bibinfo {author}
		{\bibnamefont {Goldberger}, \bibfnamefont {J.~E.}}, \bibinfo {author}
		{\bibnamefont {Yan}, \bibfnamefont {J.}}, \bibinfo {author} {\bibnamefont
			{Mandrus}, \bibfnamefont {D.~G.}}, \bibinfo {author} {\bibnamefont {Hwang},
			\bibfnamefont {J.}}, \bibinfo {author} {\bibnamefont {Cheng}, \bibfnamefont
			{R.}}, \bibinfo {author} {\bibnamefont {Katoch}, \bibfnamefont {J.}}, \ and\
		\bibinfo {author} {\bibnamefont {Singh}, \bibfnamefont {S.}},\ }\href
	{\doibase 10.1038/s41563-022-01275-5} {\bibfield  {journal} {\bibinfo
			{journal} {Nature Materials}\ }\textbf {\bibinfo {volume} {21}},\ \bibinfo
		{pages} {1029} (\bibinfo {year} {2022})},\ \Eprint
	{http://arxiv.org/abs/2012.12388} {arXiv:2012.12388} \BibitemShut {NoStop}%
	\bibitem [{\citenamefont {Kim}\ \emph {et~al.}(2013)\citenamefont {Kim},
		\citenamefont {Sinha}, \citenamefont {Hayashi}, \citenamefont {Yamanouchi},
		\citenamefont {Fukami}, \citenamefont {Suzuki}, \citenamefont {Mitani},\ and\
		\citenamefont {Ohno}}]{Kim2013}%
	\BibitemOpen
	\bibfield  {author} {\bibinfo {author} {\bibnamefont {Kim}, \bibfnamefont
			{J.}}, \bibinfo {author} {\bibnamefont {Sinha}, \bibfnamefont {J.}}, \bibinfo
		{author} {\bibnamefont {Hayashi}, \bibfnamefont {M.}}, \bibinfo {author}
		{\bibnamefont {Yamanouchi}, \bibfnamefont {M.}}, \bibinfo {author}
		{\bibnamefont {Fukami}, \bibfnamefont {S.}}, \bibinfo {author} {\bibnamefont
			{Suzuki}, \bibfnamefont {T.}}, \bibinfo {author} {\bibnamefont {Mitani},
			\bibfnamefont {S.}}, \ and\ \bibinfo {author} {\bibnamefont {Ohno},
			\bibfnamefont {H.}},\ }\href {\doibase 10.1038/nmat3522} {\bibfield
		{journal} {\bibinfo  {journal} {Nature Materials}\ }\textbf {\bibinfo
			{volume} {12}},\ \bibinfo {pages} {240} (\bibinfo {year} {2013})}\BibitemShut
	{NoStop}%
	\bibitem [{\citenamefont {Krizakova}\ \emph {et~al.}(2022)\citenamefont
		{Krizakova}, \citenamefont {Perumkunnil}, \citenamefont {Couet},
		\citenamefont {Gambardella},\ and\ \citenamefont {Garello}}]{Krizakova2022}%
	\BibitemOpen
	\bibfield  {author} {\bibinfo {author} {\bibnamefont {Krizakova},
			\bibfnamefont {V.}}, \bibinfo {author} {\bibnamefont {Perumkunnil},
			\bibfnamefont {M.}}, \bibinfo {author} {\bibnamefont {Couet}, \bibfnamefont
			{S.}}, \bibinfo {author} {\bibnamefont {Gambardella}, \bibfnamefont {P.}}, \
		and\ \bibinfo {author} {\bibnamefont {Garello}, \bibfnamefont {K.}},\ }\href
	{\doibase 10.1016/j.jmmm.2022.169692} {\bibfield  {journal} {\bibinfo
			{journal} {Journal of Magnetism and Magnetic Materials}\ }\textbf {\bibinfo
			{volume} {562}},\ \bibinfo {pages} {169692} (\bibinfo {year}
		{2022})}\BibitemShut {NoStop}%
	\bibitem [{\citenamefont {Kurebayashi}\ \emph {et~al.}(2022)\citenamefont
		{Kurebayashi}, \citenamefont {Garcia}, \citenamefont {Khan}, \citenamefont
		{Sinova},\ and\ \citenamefont {Roche}}]{Kurebayashi2022}%
	\BibitemOpen
	\bibfield  {author} {\bibinfo {author} {\bibnamefont {Kurebayashi},
			\bibfnamefont {H.}}, \bibinfo {author} {\bibnamefont {Garcia}, \bibfnamefont
			{J.~H.}}, \bibinfo {author} {\bibnamefont {Khan}, \bibfnamefont {S.}},
		\bibinfo {author} {\bibnamefont {Sinova}, \bibfnamefont {J.}}, \ and\
		\bibinfo {author} {\bibnamefont {Roche}, \bibfnamefont {S.}},\ }\href
	{\doibase 10.1038/s42254-021-00403-5} {\bibfield  {journal} {\bibinfo
			{journal} {Nature Reviews Physics}\ }\textbf {\bibinfo {volume} {4}},\
		\bibinfo {pages} {150} (\bibinfo {year} {2022})},\ \Eprint
	{http://arxiv.org/abs/2107.03763} {arXiv:2107.03763} \BibitemShut {NoStop}%
	\bibitem [{\citenamefont {Lee}\ \emph {et~al.}(2014)\citenamefont {Lee},
		\citenamefont {Lee}, \citenamefont {Cho}, \citenamefont {Choi}, \citenamefont
		{You}, \citenamefont {Jung}, \citenamefont {Bonell}, \citenamefont {Shiota},
		\citenamefont {Miwa},\ and\ \citenamefont {Suzuki}}]{Lee2014}%
	\BibitemOpen
	\bibfield  {author} {\bibinfo {author} {\bibnamefont {Lee}, \bibfnamefont
			{H.-R.}}, \bibinfo {author} {\bibnamefont {Lee}, \bibfnamefont {K.}},
		\bibinfo {author} {\bibnamefont {Cho}, \bibfnamefont {J.}}, \bibinfo {author}
		{\bibnamefont {Choi}, \bibfnamefont {Y.-H.}}, \bibinfo {author} {\bibnamefont
			{You}, \bibfnamefont {C.-Y.}}, \bibinfo {author} {\bibnamefont {Jung},
			\bibfnamefont {M.-H.}}, \bibinfo {author} {\bibnamefont {Bonell},
			\bibfnamefont {F.}}, \bibinfo {author} {\bibnamefont {Shiota}, \bibfnamefont
			{Y.}}, \bibinfo {author} {\bibnamefont {Miwa}, \bibfnamefont {S.}}, \ and\
		\bibinfo {author} {\bibnamefont {Suzuki}, \bibfnamefont {Y.}},\ }\href
	{\doibase 10.1038/srep06548} {\bibfield  {journal} {\bibinfo  {journal}
			{Scientific Reports}\ }\textbf {\bibinfo {volume} {4}},\ \bibinfo {pages}
		{6548} (\bibinfo {year} {2014})}\BibitemShut {NoStop}%
	\bibitem [{\citenamefont {Li}\ \emph {et~al.}(2012)\citenamefont {Li},
		\citenamefont {Zhang}, \citenamefont {Yap}, \citenamefont {Tay},
		\citenamefont {Edwin}, \citenamefont {Olivier},\ and\ \citenamefont
		{Baillargeat}}]{Li2012}%
	\BibitemOpen
	\bibfield  {author} {\bibinfo {author} {\bibnamefont {Li}, \bibfnamefont
			{H.}}, \bibinfo {author} {\bibnamefont {Zhang}, \bibfnamefont {Q.}}, \bibinfo
		{author} {\bibnamefont {Yap}, \bibfnamefont {C.~C.~R.}}, \bibinfo {author}
		{\bibnamefont {Tay}, \bibfnamefont {B.~K.}}, \bibinfo {author} {\bibnamefont
			{Edwin}, \bibfnamefont {T.~H.~T.}}, \bibinfo {author} {\bibnamefont
			{Olivier}, \bibfnamefont {A.}}, \ and\ \bibinfo {author} {\bibnamefont
			{Baillargeat}, \bibfnamefont {D.}},\ }\href {\doibase 10.1002/adfm.201102111}
	{\bibfield  {journal} {\bibinfo  {journal} {Advanced Functional Materials}\
		}\textbf {\bibinfo {volume} {22}},\ \bibinfo {pages} {1385} (\bibinfo {year}
		{2012})}\BibitemShut {NoStop}%
	\bibitem [{\citenamefont {Liu}\ \emph {et~al.}(2012)\citenamefont {Liu},
		\citenamefont {Pai}, \citenamefont {Li}, \citenamefont {Tseng}, \citenamefont
		{Ralph},\ and\ \citenamefont {Buhrman}}]{Liu2012c}%
	\BibitemOpen
	\bibfield  {author} {\bibinfo {author} {\bibnamefont {Liu}, \bibfnamefont
			{L.}}, \bibinfo {author} {\bibnamefont {Pai}, \bibfnamefont {C.-F.}},
		\bibinfo {author} {\bibnamefont {Li}, \bibfnamefont {Y.}}, \bibinfo {author}
		{\bibnamefont {Tseng}, \bibfnamefont {H.~W.}}, \bibinfo {author}
		{\bibnamefont {Ralph}, \bibfnamefont {D.~C.}}, \ and\ \bibinfo {author}
		{\bibnamefont {Buhrman}, \bibfnamefont {R.~a.}},\ }\href {\doibase
		10.1126/science.1218197} {\bibfield  {journal} {\bibinfo  {journal}
			{Science}\ }\textbf {\bibinfo {volume} {336}},\ \bibinfo {pages} {555}
		(\bibinfo {year} {2012})},\ \Eprint {http://arxiv.org/abs/1203.2875}
	{arXiv:1203.2875} \BibitemShut {NoStop}%
	\bibitem [{\citenamefont {Liu}\ and\ \citenamefont {Shao}(2020)}]{Liu2020b}%
	\BibitemOpen
	\bibfield  {author} {\bibinfo {author} {\bibnamefont {Liu}, \bibfnamefont
			{Y.}}\ and\ \bibinfo {author} {\bibnamefont {Shao}, \bibfnamefont {Q.}},\
	}\href {\doibase 10.1021/acsnano.0c04403} {\bibfield  {journal} {\bibinfo
			{journal} {ACS Nano}\ }\textbf {\bibinfo {volume} {14}},\ \bibinfo {pages}
		{9389} (\bibinfo {year} {2020})},\ \Eprint {http://arxiv.org/abs/2003.11966}
	{arXiv:2003.11966} \BibitemShut {NoStop}%
	\bibitem [{\citenamefont {Lv}\ \emph {et~al.}(2018)\citenamefont {Lv},
		\citenamefont {Jia}, \citenamefont {Wang}, \citenamefont {Lu}, \citenamefont
		{Luo}, \citenamefont {Zhang}, \citenamefont {Zeng},\ and\ \citenamefont
		{Liu}}]{Lv2018a}%
	\BibitemOpen
	\bibfield  {author} {\bibinfo {author} {\bibnamefont {Lv}, \bibfnamefont
			{W.}}, \bibinfo {author} {\bibnamefont {Jia}, \bibfnamefont {Z.}}, \bibinfo
		{author} {\bibnamefont {Wang}, \bibfnamefont {B.}}, \bibinfo {author}
		{\bibnamefont {Lu}, \bibfnamefont {Y.}}, \bibinfo {author} {\bibnamefont
			{Luo}, \bibfnamefont {X.}}, \bibinfo {author} {\bibnamefont {Zhang},
			\bibfnamefont {B.}}, \bibinfo {author} {\bibnamefont {Zeng}, \bibfnamefont
			{Z.}}, \ and\ \bibinfo {author} {\bibnamefont {Liu}, \bibfnamefont {Z.}},\
	}\href {\doibase 10.1021/acsami.7b16919} {\bibfield  {journal} {\bibinfo
			{journal} {ACS Applied Materials and Interfaces}\ }\textbf {\bibinfo {volume}
			{10}},\ \bibinfo {pages} {2843} (\bibinfo {year} {2018})}\BibitemShut
	{NoStop}%
	\bibitem [{\citenamefont {MacNeill}\ \emph
		{et~al.}(2017{\natexlab{a}})\citenamefont {MacNeill}, \citenamefont {Stiehl},
		\citenamefont {Guimaraes}, \citenamefont {Buhrman}, \citenamefont {Park},\
		and\ \citenamefont {Ralph}}]{MacNeill2017a}%
	\BibitemOpen
	\bibfield  {author} {\bibinfo {author} {\bibnamefont {MacNeill},
			\bibfnamefont {D.}}, \bibinfo {author} {\bibnamefont {Stiehl}, \bibfnamefont
			{G.~M.}}, \bibinfo {author} {\bibnamefont {Guimaraes}, \bibfnamefont
			{M.~H.}}, \bibinfo {author} {\bibnamefont {Buhrman}, \bibfnamefont {R.~A.}},
		\bibinfo {author} {\bibnamefont {Park}, \bibfnamefont {J.}}, \ and\ \bibinfo
		{author} {\bibnamefont {Ralph}, \bibfnamefont {D.~C.}},\ }\href {\doibase
		10.1038/nphys3933} {\bibfield  {journal} {\bibinfo  {journal} {Nature
				Physics}\ }\textbf {\bibinfo {volume} {13}},\ \bibinfo {pages} {300}
		(\bibinfo {year} {2017}{\natexlab{a}})},\ \Eprint
	{http://arxiv.org/abs/1605.02712} {arXiv:1605.02712} \BibitemShut {NoStop}%
	\bibitem [{\citenamefont {MacNeill}\ \emph
		{et~al.}(2017{\natexlab{b}})\citenamefont {MacNeill}, \citenamefont {Stiehl},
		\citenamefont {Guimar{\~{a}}es}, \citenamefont {Reynolds}, \citenamefont
		{Buhrman},\ and\ \citenamefont {Ralph}}]{Macneill2017c}%
	\BibitemOpen
	\bibfield  {author} {\bibinfo {author} {\bibnamefont {MacNeill},
			\bibfnamefont {D.}}, \bibinfo {author} {\bibnamefont {Stiehl}, \bibfnamefont
			{G.~M.}}, \bibinfo {author} {\bibnamefont {Guimar{\~{a}}es}, \bibfnamefont
			{M.~H.~D.}}, \bibinfo {author} {\bibnamefont {Reynolds}, \bibfnamefont
			{N.~D.}}, \bibinfo {author} {\bibnamefont {Buhrman}, \bibfnamefont {R.~A.}},
		\ and\ \bibinfo {author} {\bibnamefont {Ralph}, \bibfnamefont {D.~C.}},\
	}\href {\doibase 10.1103/PhysRevB.96.054450} {\bibfield  {journal} {\bibinfo
			{journal} {Physical Review B}\ }\textbf {\bibinfo {volume} {96}},\ \bibinfo
		{pages} {054450} (\bibinfo {year} {2017}{\natexlab{b}})},\ \Eprint
	{http://arxiv.org/abs/1707.03757} {arXiv:1707.03757} \BibitemShut {NoStop}%
	\bibitem [{\citenamefont {Manchon}\ \emph {et~al.}(2015)\citenamefont
		{Manchon}, \citenamefont {Koo}, \citenamefont {Nitta}, \citenamefont
		{Frolov},\ and\ \citenamefont {Duine}}]{Manchon2015c}%
	\BibitemOpen
	\bibfield  {author} {\bibinfo {author} {\bibnamefont {Manchon}, \bibfnamefont
			{A.}}, \bibinfo {author} {\bibnamefont {Koo}, \bibfnamefont {H.~C.}},
		\bibinfo {author} {\bibnamefont {Nitta}, \bibfnamefont {J.}}, \bibinfo
		{author} {\bibnamefont {Frolov}, \bibfnamefont {S.~M.}}, \ and\ \bibinfo
		{author} {\bibnamefont {Duine}, \bibfnamefont {R.~A.}},\ }\href {\doibase
		10.1038/nmat4360} {\bibfield  {journal} {\bibinfo  {journal} {Nature
				Materials}\ }\textbf {\bibinfo {volume} {14}},\ \bibinfo {pages} {871}
		(\bibinfo {year} {2015})},\ \Eprint {http://arxiv.org/abs/1507.02408}
	{arXiv:1507.02408} \BibitemShut {NoStop}%
	\bibitem [{\citenamefont {Manchon}\ \emph {et~al.}(2019)\citenamefont
		{Manchon}, \citenamefont {{\v{Z}}elezn{\'{y}}}, \citenamefont {Miron},
		\citenamefont {Jungwirth}, \citenamefont {Sinova}, \citenamefont {Thiaville},
		\citenamefont {Garello},\ and\ \citenamefont {Gambardella}}]{Manchon2019}%
	\BibitemOpen
	\bibfield  {author} {\bibinfo {author} {\bibnamefont {Manchon}, \bibfnamefont
			{A.}}, \bibinfo {author} {\bibnamefont {{\v{Z}}elezn{\'{y}}}, \bibfnamefont
			{J.}}, \bibinfo {author} {\bibnamefont {Miron}, \bibfnamefont {I.~M.}},
		\bibinfo {author} {\bibnamefont {Jungwirth}, \bibfnamefont {T.}}, \bibinfo
		{author} {\bibnamefont {Sinova}, \bibfnamefont {J.}}, \bibinfo {author}
		{\bibnamefont {Thiaville}, \bibfnamefont {A.}}, \bibinfo {author}
		{\bibnamefont {Garello}, \bibfnamefont {K.}}, \ and\ \bibinfo {author}
		{\bibnamefont {Gambardella}, \bibfnamefont {P.}},\ }\href {\doibase
		10.1103/RevModPhys.91.035004} {\bibfield  {journal} {\bibinfo  {journal}
			{Reviews of Modern Physics}\ }\textbf {\bibinfo {volume} {91}},\ \bibinfo
		{pages} {035004} (\bibinfo {year} {2019})},\ \Eprint
	{http://arxiv.org/abs/1801.09636} {arXiv:1801.09636} \BibitemShut {NoStop}%
	\bibitem [{\citenamefont {Neumann}\ and\ \citenamefont
		{Meinert}(2018)}]{Neumann2018a}%
	\BibitemOpen
	\bibfield  {author} {\bibinfo {author} {\bibnamefont {Neumann}, \bibfnamefont
			{L.}}\ and\ \bibinfo {author} {\bibnamefont {Meinert}, \bibfnamefont {M.}},\
	}\href {\doibase 10.1063/1.5037391} {\bibfield  {journal} {\bibinfo
			{journal} {AIP Advances}\ }\textbf {\bibinfo {volume} {8}},\ \bibinfo {pages}
		{095320} (\bibinfo {year} {2018})},\ \Eprint
	{http://arxiv.org/abs/1804.07577} {arXiv:1804.07577} \BibitemShut {NoStop}%
	\bibitem [{\citenamefont {Nguyen}\ and\ \citenamefont
		{Pai}(2021)}]{Nguyen2021a}%
	\BibitemOpen
	\bibfield  {author} {\bibinfo {author} {\bibnamefont {Nguyen}, \bibfnamefont
			{M.~H.}}\ and\ \bibinfo {author} {\bibnamefont {Pai}, \bibfnamefont
			{C.~F.}},\ }\href {\doibase 10.1063/5.0041123} {\bibfield  {journal}
		{\bibinfo  {journal} {APL Materials}\ }\textbf {\bibinfo {volume} {9}}
		(\bibinfo {year} {2021}),\ 10.1063/5.0041123}\BibitemShut {NoStop}%
	\bibitem [{\citenamefont {Nguyen}, \citenamefont {Ralph},\ and\ \citenamefont
		{Buhrman}(2016)}]{Nguyen2016a}%
	\BibitemOpen
	\bibfield  {author} {\bibinfo {author} {\bibnamefont {Nguyen}, \bibfnamefont
			{M.-h.}}, \bibinfo {author} {\bibnamefont {Ralph}, \bibfnamefont {D.~C.}}, \
		and\ \bibinfo {author} {\bibnamefont {Buhrman}, \bibfnamefont {R.~A.}},\
	}\href {\doibase 10.1103/PhysRevLett.116.126601} {\bibfield  {journal}
		{\bibinfo  {journal} {Physical Review Letters}\ }\textbf {\bibinfo {volume}
			{116}},\ \bibinfo {pages} {126601} (\bibinfo {year} {2016})}\BibitemShut
	{NoStop}%
	\bibitem [{\citenamefont {Novakov}\ \emph {et~al.}(2021)\citenamefont
		{Novakov}, \citenamefont {Jariwala}, \citenamefont {Vu}, \citenamefont
		{Kozhakhmetov}, \citenamefont {Robinson},\ and\ \citenamefont
		{Heron}}]{Novakov2021c}%
	\BibitemOpen
	\bibfield  {author} {\bibinfo {author} {\bibnamefont {Novakov}, \bibfnamefont
			{S.}}, \bibinfo {author} {\bibnamefont {Jariwala}, \bibfnamefont {B.}},
		\bibinfo {author} {\bibnamefont {Vu}, \bibfnamefont {N.~M.}}, \bibinfo
		{author} {\bibnamefont {Kozhakhmetov}, \bibfnamefont {A.}}, \bibinfo {author}
		{\bibnamefont {Robinson}, \bibfnamefont {J.~A.}}, \ and\ \bibinfo {author}
		{\bibnamefont {Heron}, \bibfnamefont {J.~T.}},\ }\href {\doibase
		10.1021/acsami.0c19266} {\bibfield  {journal} {\bibinfo  {journal} {ACS
				Applied Materials {\&} Interfaces}\ }\textbf {\bibinfo {volume} {13}},\
		\bibinfo {pages} {13744} (\bibinfo {year} {2021})}\BibitemShut {NoStop}%
	\bibitem [{\citenamefont {Pai}\ \emph {et~al.}(2012)\citenamefont {Pai},
		\citenamefont {Liu}, \citenamefont {Li}, \citenamefont {Tseng}, \citenamefont
		{Ralph},\ and\ \citenamefont {Buhrman}}]{Pai2012}%
	\BibitemOpen
	\bibfield  {author} {\bibinfo {author} {\bibnamefont {Pai}, \bibfnamefont
			{C.-F.}}, \bibinfo {author} {\bibnamefont {Liu}, \bibfnamefont {L.}},
		\bibinfo {author} {\bibnamefont {Li}, \bibfnamefont {Y.}}, \bibinfo {author}
		{\bibnamefont {Tseng}, \bibfnamefont {H.~W.}}, \bibinfo {author}
		{\bibnamefont {Ralph}, \bibfnamefont {D.~C.}}, \ and\ \bibinfo {author}
		{\bibnamefont {Buhrman}, \bibfnamefont {R.~A.}},\ }\href {\doibase
		10.1063/1.4753947} {\bibfield  {journal} {\bibinfo  {journal} {Applied
				Physics Letters}\ }\textbf {\bibinfo {volume} {101}},\ \bibinfo {pages}
		{122404} (\bibinfo {year} {2012})}\BibitemShut {NoStop}%
	\bibitem [{\citenamefont {Ramaswamy}\ \emph {et~al.}(2016)\citenamefont
		{Ramaswamy}, \citenamefont {Qiu}, \citenamefont {Dutta}, \citenamefont
		{Pollard},\ and\ \citenamefont {Yang}}]{Ramaswamy2016}%
	\BibitemOpen
	\bibfield  {author} {\bibinfo {author} {\bibnamefont {Ramaswamy},
			\bibfnamefont {R.}}, \bibinfo {author} {\bibnamefont {Qiu}, \bibfnamefont
			{X.}}, \bibinfo {author} {\bibnamefont {Dutta}, \bibfnamefont {T.}}, \bibinfo
		{author} {\bibnamefont {Pollard}, \bibfnamefont {S.~D.}}, \ and\ \bibinfo
		{author} {\bibnamefont {Yang}, \bibfnamefont {H.}},\ }\href {\doibase
		10.1063/1.4951674} {\bibfield  {journal} {\bibinfo  {journal} {Applied
				Physics Letters}\ }\textbf {\bibinfo {volume} {108}},\ \bibinfo {pages}
		{202406} (\bibinfo {year} {2016})}\BibitemShut {NoStop}%
	\bibitem [{\citenamefont {Schippers}, \citenamefont {Swagten},\ and\
		\citenamefont {Guimar{\~{a}}es}(2020)}]{Schippers2020c}%
	\BibitemOpen
	\bibfield  {author} {\bibinfo {author} {\bibnamefont {Schippers},
			\bibfnamefont {C.~F.}}, \bibinfo {author} {\bibnamefont {Swagten},
			\bibfnamefont {H.~J.~M.}}, \ and\ \bibinfo {author} {\bibnamefont
			{Guimar{\~{a}}es}, \bibfnamefont {M.~H.~D.}},\ }\href {\doibase
		10.1103/PhysRevMaterials.4.084007} {\bibfield  {journal} {\bibinfo  {journal}
			{Physical Review Materials}\ }\textbf {\bibinfo {volume} {4}},\ \bibinfo
		{pages} {084007} (\bibinfo {year} {2020})}\BibitemShut {NoStop}%
	\bibitem [{\citenamefont {Seki}\ \emph {et~al.}(2021)\citenamefont {Seki},
		\citenamefont {Lau}, \citenamefont {Iihama},\ and\ \citenamefont
		{Takanashi}}]{Seki2021a}%
	\BibitemOpen
	\bibfield  {author} {\bibinfo {author} {\bibnamefont {Seki}, \bibfnamefont
			{T.}}, \bibinfo {author} {\bibnamefont {Lau}, \bibfnamefont {Y.-C.}},
		\bibinfo {author} {\bibnamefont {Iihama}, \bibfnamefont {S.}}, \ and\
		\bibinfo {author} {\bibnamefont {Takanashi}, \bibfnamefont {K.}},\ }\href
	{\doibase 10.1103/PhysRevB.104.094430} {\bibfield  {journal} {\bibinfo
			{journal} {Physical Review B}\ }\textbf {\bibinfo {volume} {104}},\ \bibinfo
		{pages} {094430} (\bibinfo {year} {2021})}\BibitemShut {NoStop}%
	\bibitem [{\citenamefont {Shao}\ \emph {et~al.}(2021)\citenamefont {Shao},
		\citenamefont {Li}, \citenamefont {Liu}, \citenamefont {Yang}, \citenamefont
		{Fukami}, \citenamefont {Razavi}, \citenamefont {Wu}, \citenamefont {Wang},
		\citenamefont {Freimuth}, \citenamefont {Mokrousov}, \citenamefont {Stiles},
		\citenamefont {Emori}, \citenamefont {Hoffmann}, \citenamefont {Akerman},
		\citenamefont {Roy}, \citenamefont {Wang}, \citenamefont {Yang},
		\citenamefont {Garello},\ and\ \citenamefont {Zhang}}]{Shao2021}%
	\BibitemOpen
	\bibfield  {author} {\bibinfo {author} {\bibnamefont {Shao}, \bibfnamefont
			{Q.}}, \bibinfo {author} {\bibnamefont {Li}, \bibfnamefont {P.}}, \bibinfo
		{author} {\bibnamefont {Liu}, \bibfnamefont {L.}}, \bibinfo {author}
		{\bibnamefont {Yang}, \bibfnamefont {H.}}, \bibinfo {author} {\bibnamefont
			{Fukami}, \bibfnamefont {S.}}, \bibinfo {author} {\bibnamefont {Razavi},
			\bibfnamefont {A.}}, \bibinfo {author} {\bibnamefont {Wu}, \bibfnamefont
			{H.}}, \bibinfo {author} {\bibnamefont {Wang}, \bibfnamefont {K.}}, \bibinfo
		{author} {\bibnamefont {Freimuth}, \bibfnamefont {F.}}, \bibinfo {author}
		{\bibnamefont {Mokrousov}, \bibfnamefont {Y.}}, \bibinfo {author}
		{\bibnamefont {Stiles}, \bibfnamefont {M.~D.}}, \bibinfo {author}
		{\bibnamefont {Emori}, \bibfnamefont {S.}}, \bibinfo {author} {\bibnamefont
			{Hoffmann}, \bibfnamefont {A.}}, \bibinfo {author} {\bibnamefont {Akerman},
			\bibfnamefont {J.}}, \bibinfo {author} {\bibnamefont {Roy}, \bibfnamefont
			{K.}}, \bibinfo {author} {\bibnamefont {Wang}, \bibfnamefont {J.-P.}},
		\bibinfo {author} {\bibnamefont {Yang}, \bibfnamefont {S.-H.}}, \bibinfo
		{author} {\bibnamefont {Garello}, \bibfnamefont {K.}}, \ and\ \bibinfo
		{author} {\bibnamefont {Zhang}, \bibfnamefont {W.}},\ }\href {\doibase
		10.1109/TMAG.2021.3078583} {\bibfield  {journal} {\bibinfo  {journal} {IEEE
				Transactions on Magnetics}\ }\textbf {\bibinfo {volume} {57}},\ \bibinfo
		{pages} {1} (\bibinfo {year} {2021})}\BibitemShut {NoStop}%
	\bibitem [{\citenamefont {Shao}\ \emph {et~al.}(2016)\citenamefont {Shao},
		\citenamefont {Yu}, \citenamefont {Lan}, \citenamefont {Shi}, \citenamefont
		{Li}, \citenamefont {Zheng}, \citenamefont {Zhu}, \citenamefont {Li},
		\citenamefont {Amiri},\ and\ \citenamefont {Wang}}]{Shao2016e}%
	\BibitemOpen
	\bibfield  {author} {\bibinfo {author} {\bibnamefont {Shao}, \bibfnamefont
			{Q.}}, \bibinfo {author} {\bibnamefont {Yu}, \bibfnamefont {G.}}, \bibinfo
		{author} {\bibnamefont {Lan}, \bibfnamefont {Y.-W.~W.}}, \bibinfo {author}
		{\bibnamefont {Shi}, \bibfnamefont {Y.}}, \bibinfo {author} {\bibnamefont
			{Li}, \bibfnamefont {M.-Y.~Y.}}, \bibinfo {author} {\bibnamefont {Zheng},
			\bibfnamefont {C.}}, \bibinfo {author} {\bibnamefont {Zhu}, \bibfnamefont
			{X.}}, \bibinfo {author} {\bibnamefont {Li}, \bibfnamefont {L.-J.~J.}},
		\bibinfo {author} {\bibnamefont {Amiri}, \bibfnamefont {P.~K.}}, \ and\
		\bibinfo {author} {\bibnamefont {Wang}, \bibfnamefont {K.~L.}},\ }\href
	{\doibase 10.1021/acs.nanolett.6b03300} {\bibfield  {journal} {\bibinfo
			{journal} {Nano Letters}\ }\textbf {\bibinfo {volume} {16}},\ \bibinfo
		{pages} {7514−7520} (\bibinfo {year} {2016})}\BibitemShut {NoStop}%
	\bibitem [{\citenamefont {Shi}\ \emph {et~al.}(2019)\citenamefont {Shi},
		\citenamefont {Liang}, \citenamefont {Zhu}, \citenamefont {Cai},
		\citenamefont {Pollard}, \citenamefont {Wang}, \citenamefont {Wang},
		\citenamefont {Wang}, \citenamefont {He}, \citenamefont {Yu}, \citenamefont
		{Eda}, \citenamefont {Liang},\ and\ \citenamefont {Yang}}]{Shi2019a}%
	\BibitemOpen
	\bibfield  {author} {\bibinfo {author} {\bibnamefont {Shi}, \bibfnamefont
			{S.}}, \bibinfo {author} {\bibnamefont {Liang}, \bibfnamefont {S.}}, \bibinfo
		{author} {\bibnamefont {Zhu}, \bibfnamefont {Z.}}, \bibinfo {author}
		{\bibnamefont {Cai}, \bibfnamefont {K.}}, \bibinfo {author} {\bibnamefont
			{Pollard}, \bibfnamefont {S.~D.}}, \bibinfo {author} {\bibnamefont {Wang},
			\bibfnamefont {Y.}}, \bibinfo {author} {\bibnamefont {Wang}, \bibfnamefont
			{J.}}, \bibinfo {author} {\bibnamefont {Wang}, \bibfnamefont {Q.}}, \bibinfo
		{author} {\bibnamefont {He}, \bibfnamefont {P.}}, \bibinfo {author}
		{\bibnamefont {Yu}, \bibfnamefont {J.}}, \bibinfo {author} {\bibnamefont
			{Eda}, \bibfnamefont {G.}}, \bibinfo {author} {\bibnamefont {Liang},
			\bibfnamefont {G.}}, \ and\ \bibinfo {author} {\bibnamefont {Yang},
			\bibfnamefont {H.}},\ }\href {\doibase 10.1038/s41565-019-0525-8} {\bibfield
		{journal} {\bibinfo  {journal} {Nature Nanotechnology}\ }\textbf {\bibinfo
			{volume} {14}},\ \bibinfo {pages} {945} (\bibinfo {year} {2019})}\BibitemShut
	{NoStop}%
	\bibitem [{\citenamefont {Sousa}, \citenamefont {Tatara},\ and\ \citenamefont
		{Ferreira}(2020)}]{Sousa2020}%
	\BibitemOpen
	\bibfield  {author} {\bibinfo {author} {\bibnamefont {Sousa}, \bibfnamefont
			{F.}}, \bibinfo {author} {\bibnamefont {Tatara}, \bibfnamefont {G.}}, \ and\
		\bibinfo {author} {\bibnamefont {Ferreira}, \bibfnamefont {A.}},\ }\href
	{\doibase 10.1103/PhysRevResearch.2.043401} {\bibfield  {journal} {\bibinfo
			{journal} {Physical Review Research}\ }\textbf {\bibinfo {volume} {2}},\
		\bibinfo {pages} {043401} (\bibinfo {year} {2020})},\ \Eprint
	{http://arxiv.org/abs/2005.09670} {arXiv:2005.09670} \BibitemShut {NoStop}%
	\bibitem [{\citenamefont {Stiehl}\ \emph
		{et~al.}(2019{\natexlab{a}})\citenamefont {Stiehl}, \citenamefont {Li},
		\citenamefont {Gupta}, \citenamefont {Baggari}, \citenamefont {Jiang},
		\citenamefont {Xie}, \citenamefont {Kourkoutis}, \citenamefont {Mak},
		\citenamefont {Shan}, \citenamefont {Buhrman},\ and\ \citenamefont
		{Ralph}}]{Stiehl2019d}%
	\BibitemOpen
	\bibfield  {author} {\bibinfo {author} {\bibnamefont {Stiehl}, \bibfnamefont
			{G.~M.}}, \bibinfo {author} {\bibnamefont {Li}, \bibfnamefont {R.}}, \bibinfo
		{author} {\bibnamefont {Gupta}, \bibfnamefont {V.}}, \bibinfo {author}
		{\bibnamefont {Baggari}, \bibfnamefont {I.~E.}}, \bibinfo {author}
		{\bibnamefont {Jiang}, \bibfnamefont {S.}}, \bibinfo {author} {\bibnamefont
			{Xie}, \bibfnamefont {H.}}, \bibinfo {author} {\bibnamefont {Kourkoutis},
			\bibfnamefont {L.~F.}}, \bibinfo {author} {\bibnamefont {Mak}, \bibfnamefont
			{K.~F.}}, \bibinfo {author} {\bibnamefont {Shan}, \bibfnamefont {J.}},
		\bibinfo {author} {\bibnamefont {Buhrman}, \bibfnamefont {R.~A.}}, \ and\
		\bibinfo {author} {\bibnamefont {Ralph}, \bibfnamefont {D.~C.}},\ }\href
	{\doibase 10.1103/PhysRevB.100.184402} {\bibfield  {journal} {\bibinfo
			{journal} {Physical Review B}\ }\textbf {\bibinfo {volume} {100}},\ \bibinfo
		{pages} {184402} (\bibinfo {year} {2019}{\natexlab{a}})},\ \Eprint
	{http://arxiv.org/abs/1906.01068} {arXiv:1906.01068} \BibitemShut {NoStop}%
	\bibitem [{\citenamefont {Stiehl}\ \emph
		{et~al.}(2019{\natexlab{b}})\citenamefont {Stiehl}, \citenamefont {MacNeill},
		\citenamefont {Sivadas}, \citenamefont {{El Baggari}}, \citenamefont
		{Guimar{\~{a}}es}, \citenamefont {Reynolds}, \citenamefont {Kourkoutis},
		\citenamefont {Fennie}, \citenamefont {Buhrman},\ and\ \citenamefont
		{Ralph}}]{Stiehl2019c}%
	\BibitemOpen
	\bibfield  {author} {\bibinfo {author} {\bibnamefont {Stiehl}, \bibfnamefont
			{G.~M.}}, \bibinfo {author} {\bibnamefont {MacNeill}, \bibfnamefont {D.}},
		\bibinfo {author} {\bibnamefont {Sivadas}, \bibfnamefont {N.}}, \bibinfo
		{author} {\bibnamefont {{El Baggari}}, \bibfnamefont {I.}}, \bibinfo {author}
		{\bibnamefont {Guimar{\~{a}}es}, \bibfnamefont {M.~H.}}, \bibinfo {author}
		{\bibnamefont {Reynolds}, \bibfnamefont {N.~D.}}, \bibinfo {author}
		{\bibnamefont {Kourkoutis}, \bibfnamefont {L.~F.}}, \bibinfo {author}
		{\bibnamefont {Fennie}, \bibfnamefont {C.~J.}}, \bibinfo {author}
		{\bibnamefont {Buhrman}, \bibfnamefont {R.~A.}}, \ and\ \bibinfo {author}
		{\bibnamefont {Ralph}, \bibfnamefont {D.~C.}},\ }\href {\doibase
		10.1021/acsnano.8b09663} {\bibfield  {journal} {\bibinfo  {journal} {ACS
				Nano}\ }\textbf {\bibinfo {volume} {13}},\ \bibinfo {pages} {2599} (\bibinfo
		{year} {2019}{\natexlab{b}})},\ \Eprint {http://arxiv.org/abs/1901.08908}
	{arXiv:1901.08908} \BibitemShut {NoStop}%
	\bibitem [{\citenamefont {Tanaka}\ \emph {et~al.}(2008)\citenamefont {Tanaka},
		\citenamefont {Kontani}, \citenamefont {Naito}, \citenamefont {Naito},
		\citenamefont {Hirashima}, \citenamefont {Yamada},\ and\ \citenamefont
		{Inoue}}]{Tanaka2008}%
	\BibitemOpen
	\bibfield  {author} {\bibinfo {author} {\bibnamefont {Tanaka}, \bibfnamefont
			{T.}}, \bibinfo {author} {\bibnamefont {Kontani}, \bibfnamefont {H.}},
		\bibinfo {author} {\bibnamefont {Naito}, \bibfnamefont {M.}}, \bibinfo
		{author} {\bibnamefont {Naito}, \bibfnamefont {T.}}, \bibinfo {author}
		{\bibnamefont {Hirashima}, \bibfnamefont {D.~S.}}, \bibinfo {author}
		{\bibnamefont {Yamada}, \bibfnamefont {K.}}, \ and\ \bibinfo {author}
		{\bibnamefont {Inoue}, \bibfnamefont {J.}},\ }\href {\doibase
		10.1103/PhysRevB.77.165117} {\bibfield  {journal} {\bibinfo  {journal}
			{Physical Review B}\ }\textbf {\bibinfo {volume} {77}},\ \bibinfo {pages}
		{165117} (\bibinfo {year} {2008})},\ \Eprint {http://arxiv.org/abs/0711.1263}
	{arXiv:0711.1263} \BibitemShut {NoStop}%
	\bibitem [{\citenamefont {Torrejon}\ \emph {et~al.}(2014)\citenamefont
		{Torrejon}, \citenamefont {Kim}, \citenamefont {Sinha}, \citenamefont
		{Mitani}, \citenamefont {Hayashi}, \citenamefont {Yamanouchi},\ and\
		\citenamefont {Ohno}}]{Torrejon2014}%
	\BibitemOpen
	\bibfield  {author} {\bibinfo {author} {\bibnamefont {Torrejon},
			\bibfnamefont {J.}}, \bibinfo {author} {\bibnamefont {Kim}, \bibfnamefont
			{J.}}, \bibinfo {author} {\bibnamefont {Sinha}, \bibfnamefont {J.}}, \bibinfo
		{author} {\bibnamefont {Mitani}, \bibfnamefont {S.}}, \bibinfo {author}
		{\bibnamefont {Hayashi}, \bibfnamefont {M.}}, \bibinfo {author} {\bibnamefont
			{Yamanouchi}, \bibfnamefont {M.}}, \ and\ \bibinfo {author} {\bibnamefont
			{Ohno}, \bibfnamefont {H.}},\ }\href {\doibase 10.1038/ncomms5655} {\bibfield
		{journal} {\bibinfo  {journal} {Nature Communications}\ }\textbf {\bibinfo
			{volume} {5}},\ \bibinfo {pages} {4655} (\bibinfo {year} {2014})}\BibitemShut
	{NoStop}%
	\bibitem [{\citenamefont {Veneri}, \citenamefont {Perkins},\ and\ \citenamefont
		{Ferreira}(2022)}]{Veneri2022}%
	\BibitemOpen
	\bibfield  {author} {\bibinfo {author} {\bibnamefont {Veneri}, \bibfnamefont
			{A.}}, \bibinfo {author} {\bibnamefont {Perkins}, \bibfnamefont {D.~T.~S.}},
		\ and\ \bibinfo {author} {\bibnamefont {Ferreira}, \bibfnamefont {A.}},\
	}\href {\doibase 10.1103/PhysRevB.106.235419} {\bibfield  {journal} {\bibinfo
			{journal} {Physical Review B}\ }\textbf {\bibinfo {volume} {106}},\ \bibinfo
		{pages} {235419} (\bibinfo {year} {2022})},\ \Eprint
	{http://arxiv.org/abs/2208.07296} {arXiv:2208.07296} \BibitemShut {NoStop}%
	\bibitem [{\citenamefont {Wang}\ \emph {et~al.}(2019)\citenamefont {Wang},
		\citenamefont {Wang}, \citenamefont {Amin}, \citenamefont {Wang},
		\citenamefont {Radhakrishnan}, \citenamefont {Davidson}, \citenamefont
		{Allen}, \citenamefont {Silva}, \citenamefont {Ohldag}, \citenamefont
		{Balzar}, \citenamefont {Zink}, \citenamefont {Haney}, \citenamefont {Xiao},
		\citenamefont {Cahill}, \citenamefont {Lorenz},\ and\ \citenamefont
		{Fan}}]{Wang2019a}%
	\BibitemOpen
	\bibfield  {author} {\bibinfo {author} {\bibnamefont {Wang}, \bibfnamefont
			{W.}}, \bibinfo {author} {\bibnamefont {Wang}, \bibfnamefont {T.}}, \bibinfo
		{author} {\bibnamefont {Amin}, \bibfnamefont {V.~P.}}, \bibinfo {author}
		{\bibnamefont {Wang}, \bibfnamefont {Y.}}, \bibinfo {author} {\bibnamefont
			{Radhakrishnan}, \bibfnamefont {A.}}, \bibinfo {author} {\bibnamefont
			{Davidson}, \bibfnamefont {A.}}, \bibinfo {author} {\bibnamefont {Allen},
			\bibfnamefont {S.~R.}}, \bibinfo {author} {\bibnamefont {Silva},
			\bibfnamefont {T.~J.}}, \bibinfo {author} {\bibnamefont {Ohldag},
			\bibfnamefont {H.}}, \bibinfo {author} {\bibnamefont {Balzar}, \bibfnamefont
			{D.}}, \bibinfo {author} {\bibnamefont {Zink}, \bibfnamefont {B.~L.}},
		\bibinfo {author} {\bibnamefont {Haney}, \bibfnamefont {P.~M.}}, \bibinfo
		{author} {\bibnamefont {Xiao}, \bibfnamefont {J.~Q.}}, \bibinfo {author}
		{\bibnamefont {Cahill}, \bibfnamefont {D.~G.}}, \bibinfo {author}
		{\bibnamefont {Lorenz}, \bibfnamefont {V.~O.}}, \ and\ \bibinfo {author}
		{\bibnamefont {Fan}, \bibfnamefont {X.}},\ }\href {\doibase
		10.1038/s41565-019-0504-0} {\bibfield  {journal} {\bibinfo  {journal} {Nature
				Nanotechnology}\ }\textbf {\bibinfo {volume} {14}},\ \bibinfo {pages} {819}
		(\bibinfo {year} {2019})}\BibitemShut {NoStop}%
	\bibitem [{\citenamefont {Yang}\ \emph {et~al.}(2022)\citenamefont {Yang},
		\citenamefont {Valenzuela}, \citenamefont {Chshiev}, \citenamefont {Couet},
		\citenamefont {Dieny}, \citenamefont {Dlubak}, \citenamefont {Fert},
		\citenamefont {Garello}, \citenamefont {Jamet}, \citenamefont {Jeong},
		\citenamefont {Lee}, \citenamefont {Lee}, \citenamefont {Martin},
		\citenamefont {Kar}, \citenamefont {S{\'{e}}n{\'{e}}or}, \citenamefont
		{Shin},\ and\ \citenamefont {Roche}}]{Yang2022}%
	\BibitemOpen
	\bibfield  {author} {\bibinfo {author} {\bibnamefont {Yang}, \bibfnamefont
			{H.}}, \bibinfo {author} {\bibnamefont {Valenzuela}, \bibfnamefont {S.~O.}},
		\bibinfo {author} {\bibnamefont {Chshiev}, \bibfnamefont {M.}}, \bibinfo
		{author} {\bibnamefont {Couet}, \bibfnamefont {S.}}, \bibinfo {author}
		{\bibnamefont {Dieny}, \bibfnamefont {B.}}, \bibinfo {author} {\bibnamefont
			{Dlubak}, \bibfnamefont {B.}}, \bibinfo {author} {\bibnamefont {Fert},
			\bibfnamefont {A.}}, \bibinfo {author} {\bibnamefont {Garello}, \bibfnamefont
			{K.}}, \bibinfo {author} {\bibnamefont {Jamet}, \bibfnamefont {M.}}, \bibinfo
		{author} {\bibnamefont {Jeong}, \bibfnamefont {D.-E.}}, \bibinfo {author}
		{\bibnamefont {Lee}, \bibfnamefont {K.}}, \bibinfo {author} {\bibnamefont
			{Lee}, \bibfnamefont {T.}}, \bibinfo {author} {\bibnamefont {Martin},
			\bibfnamefont {M.-B.}}, \bibinfo {author} {\bibnamefont {Kar}, \bibfnamefont
			{G.~S.}}, \bibinfo {author} {\bibnamefont {S{\'{e}}n{\'{e}}or}, \bibfnamefont
			{P.}}, \bibinfo {author} {\bibnamefont {Shin}, \bibfnamefont {H.-J.}}, \ and\
		\bibinfo {author} {\bibnamefont {Roche}, \bibfnamefont {S.}},\ }\href
	{\doibase 10.1038/s41586-022-04768-0} {\bibfield  {journal} {\bibinfo
			{journal} {Nature}\ }\textbf {\bibinfo {volume} {606}},\ \bibinfo {pages}
		{663} (\bibinfo {year} {2022})}\BibitemShut {NoStop}%
	\bibitem [{\citenamefont {Yu}\ \emph {et~al.}(2014)\citenamefont {Yu},
		\citenamefont {Upadhyaya}, \citenamefont {Fan}, \citenamefont {Alzate},
		\citenamefont {Jiang}, \citenamefont {Wong}, \citenamefont {Takei},
		\citenamefont {Bender}, \citenamefont {Chang}, \citenamefont {Jiang},
		\citenamefont {Lang}, \citenamefont {Tang}, \citenamefont {Wang},
		\citenamefont {Tserkovnyak}, \citenamefont {Amiri},\ and\ \citenamefont
		{Wang}}]{Yu2014b}%
	\BibitemOpen
	\bibfield  {author} {\bibinfo {author} {\bibnamefont {Yu}, \bibfnamefont
			{G.}}, \bibinfo {author} {\bibnamefont {Upadhyaya}, \bibfnamefont {P.}},
		\bibinfo {author} {\bibnamefont {Fan}, \bibfnamefont {Y.}}, \bibinfo {author}
		{\bibnamefont {Alzate}, \bibfnamefont {J.~G.}}, \bibinfo {author}
		{\bibnamefont {Jiang}, \bibfnamefont {W.}}, \bibinfo {author} {\bibnamefont
			{Wong}, \bibfnamefont {K.~L.}}, \bibinfo {author} {\bibnamefont {Takei},
			\bibfnamefont {S.}}, \bibinfo {author} {\bibnamefont {Bender}, \bibfnamefont
			{S.~A.}}, \bibinfo {author} {\bibnamefont {Chang}, \bibfnamefont {L.~T.}},
		\bibinfo {author} {\bibnamefont {Jiang}, \bibfnamefont {Y.}}, \bibinfo
		{author} {\bibnamefont {Lang}, \bibfnamefont {M.}}, \bibinfo {author}
		{\bibnamefont {Tang}, \bibfnamefont {J.}}, \bibinfo {author} {\bibnamefont
			{Wang}, \bibfnamefont {Y.}}, \bibinfo {author} {\bibnamefont {Tserkovnyak},
			\bibfnamefont {Y.}}, \bibinfo {author} {\bibnamefont {Amiri}, \bibfnamefont
			{P.~K.}}, \ and\ \bibinfo {author} {\bibnamefont {Wang}, \bibfnamefont
			{K.~L.}},\ }\href {\doibase 10.1038/nnano.2014.94} {\bibfield  {journal}
		{\bibinfo  {journal} {Nature Nanotechnology}\ }\textbf {\bibinfo {volume}
			{9}},\ \bibinfo {pages} {548} (\bibinfo {year} {2014})}\BibitemShut {NoStop}%
	\bibitem [{\citenamefont {Zhang}\ \emph {et~al.}(2016)\citenamefont {Zhang},
		\citenamefont {Sklenar}, \citenamefont {Hsu}, \citenamefont {Jiang},
		\citenamefont {Jungfleisch}, \citenamefont {Xiao}, \citenamefont {Fradin},
		\citenamefont {Liu}, \citenamefont {Pearson}, \citenamefont {Ketterson},
		\citenamefont {Yang},\ and\ \citenamefont {Hoffmann}}]{Zhang2016a}%
	\BibitemOpen
	\bibfield  {author} {\bibinfo {author} {\bibnamefont {Zhang}, \bibfnamefont
			{W.}}, \bibinfo {author} {\bibnamefont {Sklenar}, \bibfnamefont {J.}},
		\bibinfo {author} {\bibnamefont {Hsu}, \bibfnamefont {B.}}, \bibinfo {author}
		{\bibnamefont {Jiang}, \bibfnamefont {W.}}, \bibinfo {author} {\bibnamefont
			{Jungfleisch}, \bibfnamefont {M.~B.}}, \bibinfo {author} {\bibnamefont
			{Xiao}, \bibfnamefont {J.}}, \bibinfo {author} {\bibnamefont {Fradin},
			\bibfnamefont {F.~Y.}}, \bibinfo {author} {\bibnamefont {Liu}, \bibfnamefont
			{Y.}}, \bibinfo {author} {\bibnamefont {Pearson}, \bibfnamefont {J.~E.}},
		\bibinfo {author} {\bibnamefont {Ketterson}, \bibfnamefont {J.~B.}}, \bibinfo
		{author} {\bibnamefont {Yang}, \bibfnamefont {Z.}}, \ and\ \bibinfo {author}
		{\bibnamefont {Hoffmann}, \bibfnamefont {A.}},\ }\href {\doibase
		10.1063/1.4943076} {\bibfield  {journal} {\bibinfo  {journal} {APL
				Materials}\ }\textbf {\bibinfo {volume} {4}} (\bibinfo {year} {2016}),\
		10.1063/1.4943076}\BibitemShut {NoStop}%
\end{thebibliography}
%

\end{document}